\documentclass[]{jfm}

\usepackage{newtxtext}
\usepackage{newtxmath}
\usepackage{natbib}
\usepackage{hyperref}
\hypersetup{
    colorlinks = true,
    urlcolor   = blue,
    citecolor  = black,
}
\newcommand{\RomanNumeralCaps}[1]
\linenumbers

\usepackage{subcaption}
\usepackage{amsmath,amssymb,amsfonts}
\usepackage{graphicx,wrapfig}
\usepackage{babel}
\usepackage{xcolor}
\usepackage{gensymb}
\usepackage{mathtools}
\usepackage{enumitem}
\usepackage[group-separator={,}]{siunitx}

\usepackage{booktabs}

\newcommand{\vect}[1]{\boldsymbol{#1}}
\newcommand{\mat}[1]{\mathsfbi{#1}}
\newcommand{\vecv}{\ensuremath{\vect{v}}}
\newcommand{\vecq}{\ensuremath{\vect{q}}}
\newcommand{\vecf}{\ensuremath{\vect{f}}}
\newcommand{\vecn}{\ensuremath{\vect{n}}}

\newcommand{\Hinf}{\ensuremath{{\mathcal{H}\scriptscriptstyle\infty}}}
\newcommand{\RHinf}{\ensuremath{\mathcal{R}\mathcal{H}_\infty}}
\newcommand{\Y}{\ensuremath{\vect{Y}}}
\newcommand{\U}{\ensuremath{\vect{U}}}

\newcommand{\Ep}{\mathbb{E}}
\newcommand{\Ek}{\mathcal{E}}
\newcommand{\BT}{\mathcal{B}_T}
\newcommand{\jw}{\jmath\omega}
\newcommand{\kwu}{k \omega_u}
\newcommand{\jwu}{\jmath \kwu}
\newcommand{\rms}{\ensuremath{\textsc{rms}}}
\newcommand{\Gb}{\ensuremath{G_b}}
\newcommand{\FB}{\ensuremath{\mathcal{F}_b}}

\newcommand{\rad}{\text{rad}}

\newcommand{\ord}[1]{\partial\degree\!{#1}}
\newcommand{\norm}[1]{\left\lVert#1\right\rVert}

\renewcommand{\d}[1]{\ensuremath{\operatorname{d}\!{#1}}}
\newcommand{\statespace}[4]{ \left[ \begin{array}{c|c} \mat{#1} & \mat{#2} \\ \midrule \mat{#3} & \mat{#4} \end{array}\right] } % with booktabs package

\DeclareMathOperator{\Var}{{Var}}

\usepackage[ruled,vlined,linesnumbered]{algorithm2e}
\DontPrintSemicolon
\SetKw{KwBreak}{break}

\newcommand{\paragraph}[1]{\emph{#1}}

\title{
Data-driven stabilization of an oscillating flow with LTI controllers
}

\author{William Jussiau\aff{1}
  \corresp{\email{william.jussiau@gmail.com}},
 Colin Leclercq\aff{2},
 Fabrice Demourant\aff{1}
 \and Pierre Apkarian\aff{1}}

\affiliation{
\aff{1} ONERA/DTIS, Université de Toulouse, Toulouse, France
\aff{2} ONERA/DAAA, Institut Polytechnique de Paris, Meudon, France
}

\begin{document}
\maketitle

%%%%%%%%%%%%%%%%%%%%%%%%%%%%%%
% Abstract
\begin{abstract}
This paper presents advances towards the data-based control of periodic oscillator flows, from their fully-developed regime to their equilibrium stabilized in closed-loop, with linear time-invariant (LTI) controllers. The proposed approach directly builds upon \cite{leclercq2019} and provides several improvements for an efficient online implementation, aimed at being applicable in experiments. First, we use input-output data to construct an LTI mean transfer functions of the flow. The model is subsequently used for the design of an LTI controller with Linear Quadratic Gaussian (LQG) synthesis, that is practical to automate online. Then, using the controller in a feedback loop, the flow shifts in phase space and oscillations are damped. The procedure is repeated until equilibrium is reached, by stacking controllers and performing balanced truncation to deal with the increasing order of the compound controller. 
In this article, we illustrate the method on the classic flow past a cylinder at Reynolds number $\Rey=100$. Care has been taken such that the method may be fully automated and hopefully used as a valuable tool in a forthcoming experiment.
\end{abstract}

\begin{keywords}
Authors should not enter keywords on the manuscript, as these must be chosen by the author during the online submission process and will then be added during the typesetting process (see \href{https://www.cambridge.org/core/journals/journal-of-fluid-mechanics/information/list-of-keywords}{Keyword PDF} for the full list).  Other classifications will be added at the same time.
\end{keywords}

{\bf MSC Codes }  {\it(Optional)} Please enter your MSC Codes here

%%%%%%%%%%%%%%%%%%%%%%%%%%%%%%%%%%%%%%
\section{Introduction}\label{sec_introduction}
For decades, understanding and controlling fluid flows has proven to be a considerable challenge due to its inherent complexity and potential impact, attracting attention from technological and academic research \citep{brunton2015}. 
In engineering, applications span several domains, from drag reduction in transport, to lift increase or acoustic noise reduction for aircraft, or mixing enhancement in chemical processes. 
Earlier approaches mainly used passive control (i.e. without exogenous input), while active control of flows is today an ongoing and fruitful research topic \citep{schmid2016,viqueratreview}. Active flow control can be itself separated into two categories: open-loop control and closed-loop control. 
While open-loop control proves to be effective (e.g. with harmonic signals in \cite{bergmann2005}), it usually requires large and sustained control inputs to drive the system to a more beneficial regime.  % sipp2012open
On the other hand, closed-loop control aims at modifying the intrinsic dynamics of the system, therefore usually requiring less energy, at the cost of an increased complexity in the design and implementation phases.

\begin{figure}
  \centering
\includegraphics[width=0.65\textwidth, trim={0.5cm 2cm 1.5cm 1cm}, clip]{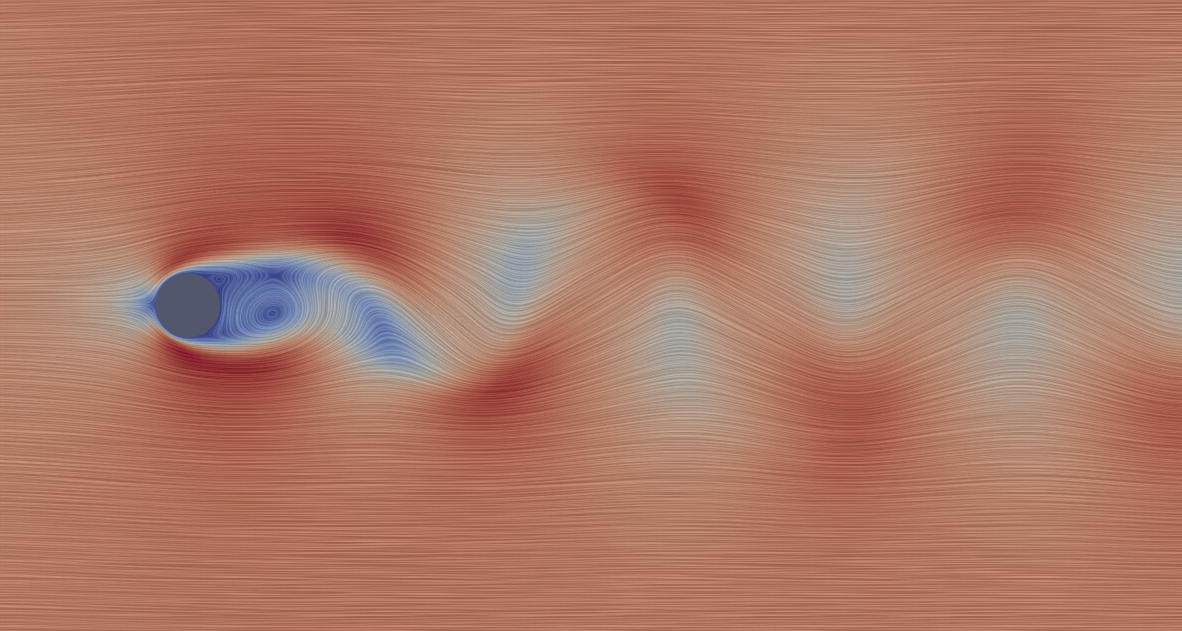}
\caption{Streamlines of a snapshot of the incompressible flow past a two-dimensional cylinder at Reynolds number $\Rey=100$. Colored by velocity magnitude.}
\label{fig_intro_cylinder}
\end{figure}

In this article, we are interested in the feedback control of laminar \emph{oscillator flows} \citep{schmid2016}.  %sipp2016
They form a particular class of flows that are linearly unstable, dominated by a nonlinear regime of self-sustained oscillations and mostly insensitive to upstream perturbations. 
The most well-known example of oscillator flow is probably the flow past a cylinder in two dimensions \citep{barkley2006}, displaying a wake of alternating vortices known as the \emph{Von Kármán vortex street} (see figure \ref{fig_intro_cylinder}). This category of flows generally exhibits two equilibria or more: an unstable fixed-point (referred to as the \emph{base flow}) and an unsteady attractor. In this application, the objective is to drive the flow from an initial state lying on the attractor, to the base flow stabilized in closed-loop. It is notable that open-loop control strategies cannot stabilize the equilibrium, while closed-loop strategies can be designed as such.

From a control perspective, closed-loop strategies developed in the literature may be categorized based on the flow regime they aim to address.
The first category of approaches focuses on preventing the growth of linear instabilities from a neighborhood of the stationary equilibrium, while the second category of approaches tackles the reduction of oscillations in the fully-developed nonlinear regime of self-sustained oscillations.
For the control law design, the main difficulty is due to the Navier-Stokes equations that are both nonlinear and infinite-dimensional.
Often, the choice of the regime of interest naturally induces a structure for the control policy (e.g. linear or nonlinear, static or dynamic...) and/or a controller design method. 
In the following, we propose to sort control approaches depending on the flow regime addressed, and we showcase some of the methods used to circumvent the difficulty posed by the high-dimensionality and nonlinearity of the controlled system.

\subsection{Control in a neighborhood of the equilibrium}
The historical approaches to oscillator flow control aim at efficiently counteracting the linear instabilities, whose development from equilibrium is eventually responsible of self-sustained oscillations. 
To do so, one may linearize the equations about the unstable equilibrium and work in the linear regime, in a neighborhood of the equilibrium.
For applying linear synthesis techniques to the inherently high-dimensional system, a common workaround is the use of a linear Reduced-Order Model (ROM) that captures the most essential features of the flow around a setpoint. Linear ROMs may be established with a wide range of techniques, such as Galerkin projection of the governing equations \citep{barbagallo2009,barbagallo2011input,weller2009feedback}, system identification from time data \citep{illingworth2011,flinois2016feedback} or frequency data \citep{jin2020feedback}, $\Hinf$ balanced truncation \citep{bennerhinf} or low-order conceptual physical modeling \citep{illingworth2012}.
Using linear ROMs, linear control techniques of various complexity may be applied: the control techniques range from proportional control \citep{weller2009feedback} and LQG \citep{barbagallo2009,barbagallo2011input,illingworth2012,illingworth2011} to $\Hinf$ synthesis or loop-shaping \citep{bennerhinf,flinois2016feedback,jin2020feedback}. In an attempt to address some shortcomings of the linearized approaches, a Linear Parameter Varying (LPV) approach was proposed in \cite{heiland2023} for both low-order modeling and control; it shows an expanded basin of attraction of control strategies around the equilibrium for a weakly supercritical flow.

These approaches are mainly restricted by the region of validity of the low-dimensional linearized model of the flow, and the assumed knowledge of the equations. In particular, these approaches are only satisfying in the vicinity of the equilibrium (or for weakly supercritical flows), but rapidly fail for strong nonlinearity when linearization about the equilibrium becomes irrelevant \citep{schmid2016}.

\subsection{Control of the fully-developed regime}
The second category of approaches tackles the reduction of oscillations in the fully-developed nonlinear regime of self-sustained oscillations. To address this regime, both data-based and model-based approaches may be suitable.

\subsubsection{Nonlinear reduced-order modeling and control}
In order to address the shortcomings of approaches using linear models, approaches were developed to handle the nonlinearity in different ways, especially with low-dimension nonlinear approximations of the flow.
Linear ROMs may be extended with nonlinear terms in Galerkin projection, in order to reproduce the oscillating behavior of the flow, in e.g. \cite{rowley2005jutti,king2005,lasagna2016}. They may be consequently used for the design of various linear and nonlinear control methods such as LPV pole placement (\cite{king2014}), Model Predictive Control (MPC, \cite{king2014}), backstepping \citep{king2005}, Sliding Mode Control (SMC, \cite{king2010}), the Sum-of-Squares formulation (SOS, \cite{lasagna2016,lasagna2017}) or more physically-based solutions \citep{rowley2005jutti,king2003}. 
These studies show that nonlinear ROMs may be used for the design of control methods and provide satisfying performance on the high-dimensional nonlinear system. 
They however might require strong model assumptions for building a nonlinear ROM (e.g. modes computation, full-field information...), which may hinder their applicability in experiments with localized measurements, noise or model mismatch.

\subsubsection{Approaches using the Koopman operator} %
More recent approaches use input-output data to directly build a \emph{surrogate} nonlinear model of the flow, and use it for control with the MPC framework, leveraging the prediction power of cheap surrogate models.
Rooted in Koopman operator theory, \cite{kordamezic,arbabi2018} develop a framework for computing a linear representation of a dynamical system, in a user-chosen lifted coordinate space. Using a moderate-dimension flow, they show the possibility of replacing full-state measurement by sparse measurements with delay embedding. A similar approach is used in \cite{morton2018} with full-state measurement but the space of observables is learned with an encoder Neural Network (NN), which is illustrated for a weakly supercritical cylinder wake flow. 
In the same idea, using Koopman operator theory, \cite{peitz2020summary} summarizes two findings introduced in two previous papers: in \cite{peitz2019switch}, the flow under any actuation signal is modeled using autonomous systems with constant input (derived from data), and the control problem is turned into a switching problem; in \cite{peitz2020interp}, a bilinear model is interpolated from said autonomous, constant-input flow models and the control problem is solved on the bilinear model.
In a related work, \cite{otto2022} directly build a bilinear model using delayed sparse observables and apply said model to the fluidic pinball, even providing decent performance at off-design Reynolds number.
In \cite{kerswell}, it is shown that a single Koopman expansion might not be able to reproduce the behavior of the flow around two distinct invariant solutions (namely, the stationary equilibrium and the attractor), which may underline the necessity of building several such models for efficient model-based control.
Not explicitly linked to the Koopman operator, \cite{bieker2019} model the actuated flow as a black-box, in a latent space with a Recurrent Neural Network (RNN) that may be updated online, and demonstrate the possibility to efficiently address complex flows such as the chaotic fluidic pinball, by only using a small number of localized measurements, provided large amounts of training data are available.

\subsubsection{Interacting with the high-dimensional nonlinear system}
On the other side of the spectrum, some techniques try to address the control of high-dimensional nonlinear systems with direct interaction with the system itself.
On one side, some of these approaches use tools and controller structures from linear control theory, such as with Proportional Integral Derivative (PID) control \citep{park1994,choi2018,yun2022}, or structured $\Hinf$ control \citep{jussiau2022learning}, in order to design controllers by direct interaction with the full-size nonlinearity, either heuristically \citep{park1994,yun2022} or with the help of optimization \citep{choi2018,jussiau2022learning}.
Furthermore, a significant body of literature use nonlinear model-free control laws, using information conveyed by the full-order nonlinear system. Examples of such approaches are found in \cite{cornejo2021} and \cite{cornejo2022}, where nonlinear control laws consisting of nested operations ($+,\times,\cos,\dotsc$) are built with optimization, for applications to the fluidic pinball. Naturally, the Reinforcement Learning (RL) framework has also proved its efficiency at discovering control policies (see \cite{viqueratreview,viqueratreviewupdate} for a review of approaches, or \cite{paris,rabault,ghraieb2021} for illustrations).
A recent study in RL \citep{rigas2023} demonstrates the effectiveness of including delayed measurements and past control inputs within a nonlinear control policy in discrete-time. This approach essentially transforms the control law from static to dynamic, utilizing the concept of \textit{dynamic output feedback} from control theory \citep{staticoutputfeedback}.

While the use of data makes these approaches more easily applicable in experiments, they still require large amounts of data and their training can be made more challenging by various external factors, such as convective time delays stemming from convective phenomena or partial observability (i.e. localized sensing, often requiring numerous sensors to reconstruct worthy information).

\subsection{Proposed approach}
\begin{figure}
\centering
\includegraphics[width=1\textwidth]{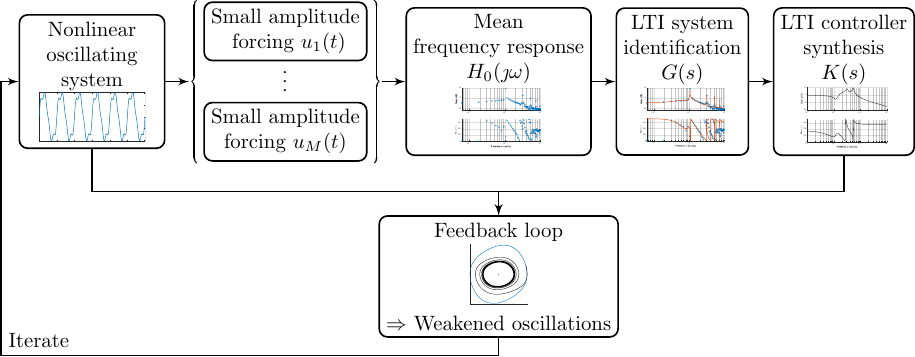}
\caption{Graphical summary of the method : data-based stabilization of an oscillating flow with LTI controllers, using the mean resolvent framework \citep{leclercq2023}.}
\label{fig_graphical_summary}
\end{figure}

In this paper, we aim at driving the system from its natural limit cycle to its equilibrium, stabilized in closed-loop, by handling the nonlinearity iteratively, in the same idea as \cite{leclercq2019} (that completely suppress oscillations on top of a cavity, by solving the nonlinear control problem as a sequence of low-order linear approximations).
The method we propose is fully data-based, but aims at being easy to design and implement, and does not require multiple sensors, extensive training nor tricky parameter tuning. It aims at handling the nonlinearity iteratively, and tackles the large dimension of the system with system identification solely from input-output data.
Using the \emph{mean resolvent} framework from \cite{leclercq2023}, we can establish an LTI model of the oscillating flow, uniquely from input-output data, which is used to design a dynamic output feedback LTI controller. While the constructed controller successfully reduces oscillations in the flow, it cannot stabilize the flow completely due to the local validity (in phase space) of the LTI model. Consequently, the flow reaches a new dynamical equilibrium characterized by a lower perturbation kinetic energy. The procedure is then iterated from this new dynamical equilibrium, until the flow is fully stabilized -- the procedure is illustrated in figure \ref{fig_graphical_summary}.

The paper is structured as follows. In section \ref{sec_method}, we present and justify the method and its associated tools in details. In section \ref{sec_results}, we demonstrate the applicability of the method on the canonical two-dimensional flow past a cylinder at $\Rey=100$, essentially reaching equilibrium with data-driven LTI controllers, and analyze the solution found. We discuss several points of the method in \S \ref{sec_discussion}. Finally, the main results are recalled and perspectives are drawn in \S \ref{sec_conclusion}.

%%%%%%%%%%%%%%%%%%%%%%%%%%%%%%%%%%%%%%
\section{Method and tools}\label{sec_method}
\subsection{Objective and overview of the method}
We consider the flow past a cylinder in two dimensions, at Reynolds number $\Rey=100$ (it is presented in more details in section \ref{sec_simulation}). This flow is an oscillator flow with an unstable equilibrium, referred to as the \emph{base flow}, and a periodic attractor, which is the regime naturally observed: the Von Kármán vortex street (figure \ref{fig_intro_cylinder}). Our objective is to drive the system from its attractor back to its equilibrium stabilized in closed-loop, using a single local sensor and a single actuator, in a fully data-driven way (i.e. without the need for knowledge of the equations, the base flow or sensor/actuator models). 

The procedure is based on the same idea as the one described in \cite{leclercq2019}. In this previous study, the oscillating flow is modeled, from an input-output viewpoint, as an LTI system enabling LTI controller design. However, as the controller is unable to completely stabilize the flow, the feedback system converges to a new attractor with lower perturbation energy, and the procedure is reiterated until the base flow is reached. More specifically, in \cite{leclercq2019}, the authors used a linearized model around the mean flow (i.e. equations linearized around the temporal average $\bar{\vecq} = \langle \vecq(t) \rangle_t$ of a statistically-steady flow), which is shown to represent features of the flow important for control \citep{liu2018resolvent}. Although the mean flow may be estimated in experiments, it remains quite impractical for applications, and the linearization performed requires significant assumptions (e.g. models for both the sensor and actuator, neglecting three-dimensional effects or working with expensive 3D equations...). 
Also, they used multi-objective structured $\Hinf$ synthesis \citep{apkarian2006nonsmooth,apkarian2014multi} for the design of low-order controllers. While this technique is very powerful and well-suited to this problem (as it can enforce e.g. controller structure, roll-off, or location of poles in the closed-loop), it is not easy to automate and often requires the control engineer perspective to be used at its maximum potential. Finally, as the controllers are being stacked onto each other during the iterations, the controller effectively operating in closed-loop has its order increasing linearly with the number of iterations. In an experiment where the procedure would likely never really converge to a steady equilibrium (due, at least, to residual incoming turbulence), this ever-increasing order could be a problem for runtime and numerical conditioning.

In this study, we tackle these three shortcomings preventing the use of the method in an automated data-based manner, summarized in table \ref{table_summary}.
First, the modeling part is done solely with input-output data, using the mean resolvent framework introduced in \cite{leclercq2023}, identified with multisine excitations \citep{schoukens1991}. Not only is the mean transfer function easier to derive and implement, but it is also better founded than the resolvent around the mean flow \citep{leclercq2023}. Second, the controller is designed with LQG synthesis, that is easy to automate. While this method is arguably less powerful and flexible than structured $\Hinf$ synthesis, it permits easy automation while maintaining some desirable properties for the controller. Third, the increasing order of stacked controllers is managed with online controller reduction with balanced truncation (denoted $\BT$ below; see \cite{moore1981,zhou1999}) and transient regimes are handled with a two-step initialization of the new controller. Overall, these advantages aim at making the method applicable in a real-life setup.
\begin{table}
\begin{center}
\begin{tabular}{p{0.2\textwidth}p{0.3\textwidth}p{0.35\textwidth}}
Step & \cite{leclercq2019} & Our proposition \\
\hline
Modeling & Linearization of the dynamics around the time-averaged flow & Identification of mean transfer function from input-output data (\S\ref{sec_identification})  \\
\hline
Control & Structured $\Hinf$  & Automated LQG (\S\ref{sec_control}) \\
\hline
Controller update & Full-order stacking & Stacking, balanced truncation and state initialization (\S\ref{sec_reduction}) \\
\end{tabular}
\caption{Rendering \cite{leclercq2019} data-based and automated.}
\label{table_summary}
\end{center}
\end{table}
The overview of the method is as follows, with $i$ the iteration index:
\begin{enumerate}[label={$\mathcal{S}$.\arabic*}]
\item~ \label{enum_simulate} Simulate flow with feedback controller $\tilde K_i(s)$. After transient regime, \\
reach new statistically-steady regime (dynamical equilibrium). 
\hfill  [\S\ref{sec_simulation}] \\
\textcolor{gray}{$\hookrightarrow$ At iteration 0, controller is null: $\tilde K_0 (s)=0$.}
 
\item~ \label{enum_model} Compute LTI ROM of the oscillating closed-loop: $G_i(s)$. 
\hfill [\S\ref{sec_identification}] \\
\textcolor{gray}{$\hookrightarrow$ With input-output data.}

\item~ \label{enum_control} Synthesize new controller $K^+_i(s)$ for the identified ROM.  
\hfill [\S\ref{sec_control}]  \\
\textcolor{gray}{$\hookrightarrow$ Automated synthesis.} 

\item~ \label{enum_stack} Stack controllers and reduce its order with the balanced truncation $\BT$: 
\\ define
 $\tilde K_{i+1}(s) = \BT(\tilde K_{i}(s) + K_i^+(s))$ to use in closed-loop. 
\hfill  [\S\ref{sec_reduction}] \\
\textcolor{gray}{$\hookrightarrow$ Reduce controller order and control input transient.}

\item~ Back to \ref{enum_simulate} and iterate until condition is met. \\
\textcolor{gray}{$\hookrightarrow$ For example, low norm of sensing signal.}

\end{enumerate}

\subsection{Notations}
\subsubsection{Notations for the iterative procedure}
\begin{figure}
\centering
\includegraphics[width=0.8\textwidth]{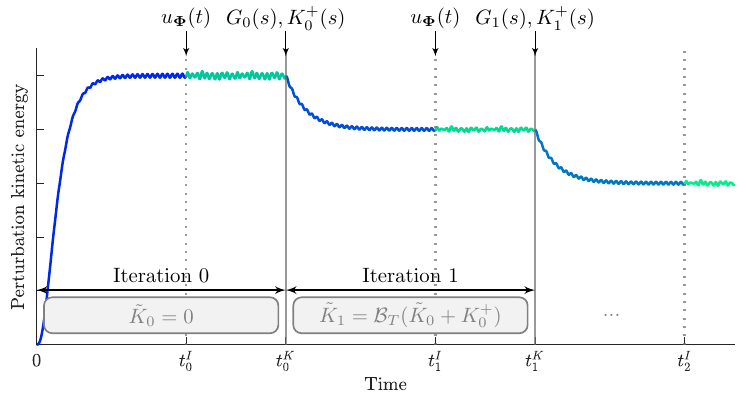}
\caption{At each iteration $i$, a time simulation is performed in closed-loop with the controller $\tilde{K}_i(s) $; then, an exogenous signal $u_{\mathbf{\vect{\Phi}}}(t)$ is injected for the identification of an LTI model $G_i(s)$, for which an LTI controller $K_i^+(s)$ is synthesized. This corresponds to the start of iteration $i+1$, where the controller in the loop is $\tilde{K}_{i+1} = \BT(\tilde{K}_i + K_i^+)$ that should drive the flow to a new dynamical equilibrium with lower perturbation kinetic energy. The process is then repeated.
}
\label{fig_graphical_summary_notations}
\end{figure}

One repetition of the identification-control process is referred to as an iteration, and they are repeated until the equilibrium is reached. The following paragraph is described graphically in figure \ref{fig_graphical_summary_notations}.
At the start of the process, numbered iteration $i=0$, the flow is simulated from its perturbed unstable equilibrium, with the feedback controller $\tilde{K}_{0}=0$. Therefore, the flow evolves towards its natural limit cycle  (\ref{enum_simulate}).
When the limit cycle is attained, at time $t_0^I$, an exogenous signal $u_{\mathbf{\vect{\Phi}}}(t)$ is injected for the identification of a model $G_0(s)$ of the oscillating flow with data from $t \in [t_0^I, t_0^K]$ (\ref{enum_model}). 
Then, a controller $K_0^+(s)$ is synthesized (\ref{enum_control}) to control $G_0(s)$. 
At time $t < t_0^K$, only the controller $K_0=0$ is in the loop. 
At time $t \geq t_0^K$, the new full-order controller is $K_1 = \tilde{K}_0+K_0^+$. After a short duration $T_{sw}$ (explained in \S\ref{sec_controller_switching}), $K_1$ is reduced with balanced truncation $\BT$ and its low-order counterpart $\tilde{K}_1=\BT(K_0+K_0^+)$ is used in its place (\ref{enum_stack}). The flow reaches a new dynamical equilibrium with lower perturbation kinetic energy. Iteration $i=1$ starts at $t=t_0^K$, and the notations are alike for the rest of the procedure.

\subsubsection{Control theory notations}
The order of an LTI plant $G$ is denoted $\ord{G} \in \mathbb{N}$. The state-space representation of a transfer $G(s)$ with associated matrices $(\mat{A}, \mat{B}, \mat{C}, \mat{D})$ is denoted:
\begin{equation} 
G=\statespace{A}{B}{C}{D},
\end{equation}  
such that the state, input and output $\vect{x}, \vect{u}, \vect{y}$ associated to this state-space realization of $G$ follow:
\begin{equation} 
\centering
\left\{
\begin{aligned}
&\dot{\vect{x}} = \mat{A}\vect{x} + \mat{B}\vect{u}, \\
& \vect{y} = \mat{C}\vect{x} + \mat{D}\vect{u}.
\end{aligned}
\right.
\end{equation} 
Note that most of the plants used in this study are single-input, single-output (SISO), i.e. with scalar $u, y$.

%%%%%%%%%%%%%%%
\subsection{Use case: flow past a cylinder at low Reynolds number} \label{sec_simulation}
\subsubsection{Configuration}
In this paper, the use case is the incompressible bidimensional flow past a cylinder, used in numerous past studies with slightly different setups and various control methods (see e.g. \cite{illingworth2016,paris} and many others). 
Here, the configuration is taken from \cite{jussiau2022learning} and some details are recalled below.
A cylinder of diameter $D$ is placed at the origin of a rectangular domain $\Omega$, equipped with the Cartesian coordinate system $(x_1, x_2)$.
All quantities are rendered nondimensional with respect to the cylinder diameter $D$, the uniform upstream velocity magnitude ${v_1}_\infty$ and the kinematic viscosity of the fluid $\nu$. The Reynolds number is defined as $\Rey=u_\infty D / \nu$, balancing convective and viscous terms. 
The domain extends as $ -15 \leq {x_1} \leq 20,  |x_2| \leq 10$.
\begin{figure}
\centering
\includegraphics[width=0.85\textwidth]{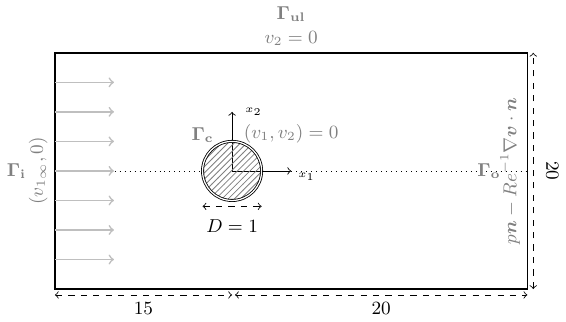}
\caption{Domain geometry for the flow past a cylinder. Dimensions are in black, while boundary conditions are in light gray. Drawing is not to scale.}
\label{fig_domain}
\end{figure}

\begin{figure}
\centering
\begin{subfigure}{.45\textwidth}
  \centering
\includegraphics[width=\textwidth, trim={3cm 4cm 1cm 4cm}, clip]{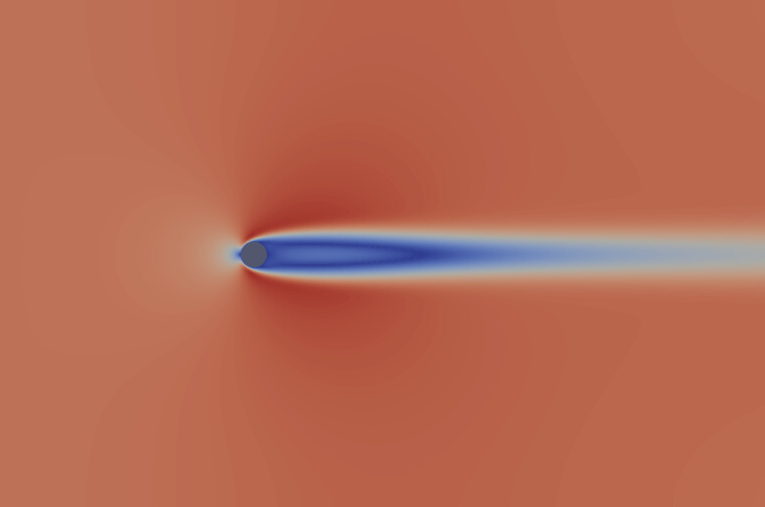} 
  \caption{}
  \label{fig_baseflow}
\end{subfigure}
\begin{subfigure}{.45\textwidth}
  \centering
\includegraphics[width=\textwidth, trim={3cm 4cm 1cm 4.1cm}, clip]{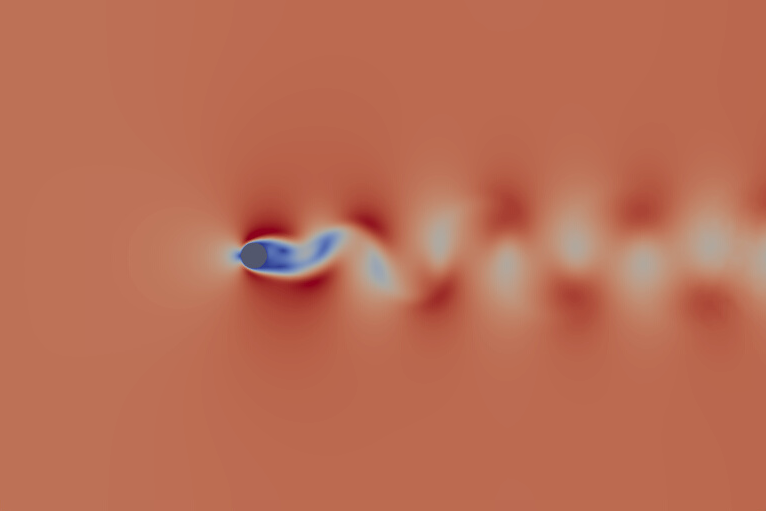} 
  \caption{}
  \label{fig_lco}
\end{subfigure}
\caption{Cylinder flow regimes (velocity magnitude) - Unstable base flow (\subref{fig_baseflow}) and snapshot of the attractor (\subref{fig_lco}). Domain is cut for clarity.}
\end{figure}

The flow is described by its velocity $\vecv = [v_1, v_2]^T$ and pressure $p$, and satisfies the incompressible Navier-Stokes (NS) equations:
\begin{equation}
\left\{
\begin{aligned} 
&  \frac{\partial \vecv}{\partial t} + (\vecv \cdot \nabla)\vecv = -\nabla p +  \frac{1}{\Rey}\nabla^2 \vecv,    \\
&  \nabla \cdot \vecv = 0.
\label{ns_eq}
\end{aligned}
\right.
\end{equation}
This dynamical system is known to undergo a supercritical Hopf bifurcation at the critical Reynolds number $\Rey_c \approx 47$ \citep{barkley2006}. Above the threshold, it displays an unstable equilibrium (here referred to as the \emph{base flow}, see figure \ref{fig_baseflow}) and a periodic attractor (i.e. a stable limit cycle, see figure \ref{fig_lco}). In this study, we set $\Rey=100$. 

\subsubsection{Boundary conditions, control and simulation setups}
\paragraph{Unactuated flow.}
A parallel flow enters from the left of the domain, directed to the right of the domain. 
The boundary conditions for the unforced flow, represented in figure \ref{fig_domain}, are detailed below: 
\begin{itemize}
\item~ On the inlet $\Gamma_i$, the fluid has parallel horizontal velocity $\vecv^{i}=({v_1}_\infty,0)$, uniform along the vertical axis $x_2$,
\item~ On the outlet $\Gamma_o$, we impose standard outflow conditions with ${p^{o}\vecn - \Rey^{-1}\nabla \vecv^{o} \cdot \vecn=0}$ where $\vecn$ is the outward-pointing vector,
\item~ On the upper and lower boundaries $\Gamma_{ul}$, that were set far from the cylinder to mitigate end effects, we impose an impermeability condition $v_2=0$,
\item~ On the surface of the cylinder $\Gamma_c$ where actuation is not active, we impose a no-slip condition with $\vecv^{c}=(0,0)$.
\end{itemize}

\paragraph{Actuation.}
In this configuration, the actuation is injection/suction of fluid at the upper and lower poles of the cylinder, in the vertical direction.
Both actuators are 10\degree-wide and impose a parabolic profile ${v_2}_p(x_1)$ to the normal velocity of the fluid, modulated by the control amplitude $u(t)$ (negative or positive). 
On the controlled boundaries, the boundary condition is $\vecv^{act}(x_1, t) = \left( 0, {v_2}_p(x_1) u(t) \right)$.
\begin{figure}
\centering
\includegraphics[width=0.45\textwidth]{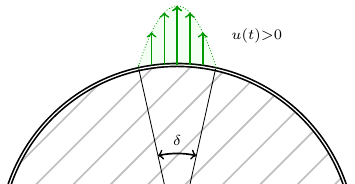}
\caption{Zoom on actuation setup.}
\label{fig_actuator}
\end{figure}
Both actuators are functioning antisymmetrically, in order to maintain an instantaneous zero-net-mass-flux. In other words, a positive actuation amplitude $u(t)>0$ corresponds to blowing from the top and suction from the bottom, and conversely with $u(t)<0$.

\paragraph{Sensing.}
It was shown in several past studies that a single-input, single-output (SISO, i.e. one actuator and one sensor) setup can be adequate for controlling the cylinder configuration \citep{jin2020feedback,flinois2016feedback,jussiau2022learning} if the sensor is positioned in order to reconstruct sufficient information, and to not suffer too much from convective time delays.
Following this trade-off, the sensor is chosen to provide the cross-stream velocity in the wake: $y(t) = v_2({x_1}\!=\!3, {x_2}\!=\!0, t)$. The sensor is assumed perfect and not corrupted by noise. Note that including a linear sensor model (e.g. limited bandwidth with a low-pass transfer function) would be seamless, as the approach is entirely based on data and only assumes linearity. 

Selecting the sensor location on the horizontal symmetry axis $x_2=0$ of the base flow yields an immediate benefit: the measurement value on the base flow can be deduced by symmetry arguments as $y_b=0$. 
This has a direct advantage, allowing the controller to operate directly on the measurement value $y(t)$ while maintaining the natural base flow as an equilibrium point of the closed-loop system. 
In the case where the sensor were placed at a location with $y_b\neq 0$, two alternatives are suggested. The first option is the controller operating on $\varepsilon(t)=y(t)-y_b$, requiring the computation of $y_b$ and $\vecq_b$, which is excluded in this data-driven approach. The second option is using a controller with zero static gain (i.e. $K(0)=0$). In this case, it could operate directly with $y(t)$, while ensuring the base flow remains an equilibrium point, and no other equilibrium with zero-input may exist, as proven in \cite{leclercq2019}.

\subsubsection{Numerical methods}
The incompressible Navier-Stokes equations in the two-dimensional domain \eqref{ns_eq} are solved in finite dimension with the Finite Element Method using the toolbox FEniCS \citep{fenics} in Python. FEniCS is widely used for solving Partial Differential Equations thanks to its very general framework and its native parallel computing capacity. 

% Mesh & FE
The unstructured mesh consists of approximately \num{25000} triangles, more densely populated in the vicinity and in the wake of the cylinder. Finite elements are chosen as Taylor-Hood $(P_2, P_2, P_1)$ 
elements, leading to a discretized descriptor system of \num{113000} states.
% Time-stepping
For time-stepping, a linear multistep method of second order is used. The nonlinear term is extrapolated with a second-order Adams-Bashforth scheme, while the viscous term is treated implicitly, making the temporal scheme semi-implicit. The time step is chosen as $\Delta t = 5\cdot 10^{-3}$ for stability and precision.
%, for stability and precision reasons. 
Each simulation is run in parallel on 24 CPU cores with distributed memory (MPI).
% Linear solver
All large-dimensional linear systems are solved with the software package MUMPS \citep{mumps}, a sparse direct solver well-suited to distributed-memory architectures and natively accessible from within FEniCS.

\subsubsection{Additional notations for characterizing the flow} \label{sec_notations_flow}
For characterizing the flow globally, we define several operations and quantities that are reused in the following. 
First, the system \eqref{ns_eq} can be written as:
\begin{equation}
\mat{E} \frac{\partial \vecq}{\partial t} = \mat{F} (\vecq),
\label{eq_ns_q}
\end{equation}
where the state variable is defined as $\vecq=[\vecv, p]^T$, $\mat{E} = \begin{bsmallmatrix} \mat{I} & 0  \\ 0 & 0 \end{bsmallmatrix}$ and the nonlinear operator $\mat{F}$ is expressed easily. 
The \emph{base flow} denoted $\vecq_b$ refers to the unique steady equilibrium of \eqref{eq_ns_q} (we recall that it is linearly unstable). The model of the flow linearized around the base flow $\vecq_b$ (or indifferently a reduced-order approximation of said model) is denoted $\Gb(s)$, and is used in the analysis of results in \S \ref{sec_results}.
We also define the semi-inner product between two fields $\vecq_1, \vecq_2$ as:
\begin{equation}
\langle \vecq_1, \vecq_2 \rangle_E = \int_{\Omega} \vecq_1^T \cdot \mat{E} \vecq_2  \, \d{\Omega}.
\end{equation}
In turn, the \emph{perturbation} kinetic energy (PKE) associated to a velocity-pressure field $\vecq$ relative to the base flow $\vecq_b$ is defined as follows:
\begin{equation}
\Ek = \frac{1}{2} \norm{\vecq - \vecq_b}_E^2.
\end{equation}
While it is only available in simulation, it will be used a posteriori to quantify the distance to the base flow, i.e. the distance to convergence. 
We also define the field $\vect{\epsilon} =  (\vecq-\vecq_b)^T \cdot \mat{E} (\vecq - \vecq_b)$, such that $\Ek = \frac{1}{2} \int_\Omega \vect{\epsilon}  \, \d{\Omega}$, that shows the distribution of the perturbation kinetic energy inside the domain $\Omega$.

Finally, when the flow is in a periodic regime (e.g. in feedback with a given controller), we define the \emph{mean flow} as the temporal average of the flow variables $\bar{\vecq} = \langle \vecq(t) \rangle_T$ over a period. It is the same quantity used in \cite{leclercq2019} for iterative identification, control and analysis of the flow, and is used in \S \ref{sec_results}, \S \ref{sec_discussion}.

%%%%%%%%%%%%%%%%%%%%%%%%%%%%%%%%%%%%%%%
\subsection{Identification of an input-output model from data leveraging the mean resolvent}\label{sec_identification}
\subsubsection{Introduction to the mean transfer function}
It is at first not obvious that the oscillating flow may be well approximated with an LTI model, moreover suitable for control purposes.
Earlier justifications were given with variations of the Dynamic Mode Decomposition (DMD) \citep{edmd,dmdc} whose theoretical link to the (linear) Koopman operator was established in \cite{kordamezicedmd}, and the models were used for control in e.g. \cite{kordamezic,kordamezicedmd,arbabi2018}.
In \cite{leclercq2023}, a new relevant model around the limit cycle is introduced, based upon observations from \cite{dahan2012,dallalonga2017,evstafyeva2017}. This model, the so-called \emph{mean resolvent}, is rooted in Floquet analysis and aims at providing the \emph{average linear response of the flow to a control input}. It is also shown to be linked to the Koopman operator \citep{leclercq2023}, and is extended to more complex attractors. This framework is briefly described in the rest of this section.

First, the \emph{linear} response of the flow refers to the response of the flow to a given control input $\vect{f}(t)$ of small amplitude, i.e. \emph{small enough} to allow linearization around the periodic deterministic unforced trajectory. Using a perturbation formulation, if the periodic unforced trajectory is denoted $\vect{Q}(t)$, the linear response to the control $\vect{f}(t)$ is $\delta \vecq(t)$, such that $\vecq(t) = \vect{Q}(t) + \delta \vecq(t)$.

Second, the \emph{average} response of the flow is considered with respect to the phase on the limit cycle when the input $\vect{f}(t)$ is injected. On a periodic attractor of pulsation $\omega$, any time instant $t$ is parametrized by a phase $\phi = \omega t \mod 2\pi \in [0, 2\pi)$, so that every point is described by its associated $\phi$. The mean resolvent $R_0(s)$ is the operator predicting, in the frequency domain, the average linear response (over $\phi$) from a given input: $\langle \delta\vecq(s; \phi) \rangle_\phi = \mat{R}_0(s) \vecf(s)$. 

Here, we focus on a SISO transfer in the flow, i.e. the transfer between a single localized actuator and a single sensor signals. The actuation signal is such that $\vect{f}(t)=\mat{B}u(t)$ and we define the measurement deviation from the limit cycle as $\delta y(t) = \mat{C} \delta \vecq(t)$, where $\mat{B}, \mat{C}$ are actuation and measurement fields, depending on the configuration. In this case, we study the SISO mean transfer function $H_0(s): u(s) \mapsto \langle \delta y(s; \phi) \rangle_\phi$, equal to $H_0(s) = \mat{C}\mat{R}_0(s)\mat{B}$. It is shown in the following, that it is possible to identify $H_0(s)$ from data, with the full measurement $y(t) = \mat{C} \vecq(t)$ (since $\delta y(t)$ is not measurable in practice).

\subsubsection{Mean frequency response}\label{sec_mean_freqresp}
The identification of $H_0(s)$ is done in two steps. First, the frequency response $H_0(\jw)$ is extracted from input-output data on a discrete grid of frequencies. 
Second, a low-order state-space model is identified from this data. The transfer function of the low-order model is denoted $G(s)$.

\paragraph{Multisine excitations.}
In order to extract the frequency response $H_0(\jw)$, we use here a particular class of input signals known as \emph{multisine excitations} \citep{schoukens1991}. As shown in \cite{leclercq2023}, any class of input signals could be used for this task (e.g. white noise, chirp, ...), with various efficiency and a potential need for ensemble averaging.
A multisine realization is a superposition of sines, depending on a random vector of independent identically-distributed uniform variables $\mathbf{\vect{\Phi}}=[\mathrm{\Phi}_1, \dotsc, \mathrm{\Phi}_N] \sim \mathcal{U}\left([0, 2\pi]^N\right)$:
\begin{equation}
u_{\mathbf{\vect{\Phi}}}(t) = \frac{2}{\sqrt N} \sum_{k=1}^{N} A_k \sin(k\omega_u t + \mathrm{\Phi}_k).
\end{equation}
The fundamental frequency of the multisine is $\omega_u$ and only harmonics $k\omega_u, 1 \leq k \leq N$ are included, each with chosen amplitude $A_k$ (normalized by $\frac{1}{2}\sqrt N$) and random phase $\mathrm{\Phi}_k$. Multisines have been chosen for their deterministic amplitude spectrum and sparse representation in the frequency domain \citep{schoukens2008robustness,schoukens2016linear}, but any other input signal could be used for the identification in this context.

\paragraph{Average of frequency responses and convergence to the mean frequency response.}
For a given input 
$u_{\mathbf{\vect{\Phi}}}(t)$, we denote 
$y_{\mathbf{\vect{\Phi}}}(t) = y(t) + \delta y_{\mathbf{\vect{\Phi}}}(t)$ the measured output, which is by definition the sum of the measurement signal of the unforced flow $y(t)$, and the forced contribution $\delta y_{\mathbf{\vect{\Phi}}}(t)$ linear with respect to $u_{\mathbf{\vect{\Phi}}}(t)$.
Following \cite{leclercq2023}, the Fourier coefficients of the input and output at the forcing frequencies $k\omega_u$ may be identified with harmonic averages:
\begin{equation}
\left\{
\begin{aligned} 
& \widehat{u}_{\mathbf{\vect{\Phi}}}(k\omega_u) = \lim_{T'\to \infty} \frac{1}{T'} \int_{0}^{T'} u_{\mathbf{\vect{\Phi}}}(t) e^{-\jwu t} \, \d t, \\
& \widehat{y}_{\mathbf{\vect{\Phi}}}(k\omega_u) = \lim_{T'\to \infty} \frac{1}{T'} \int_{0}^{T'} y_{\mathbf{\vect{\Phi}}}(t) e^{-\jwu t} \, \d t,
\end{aligned}
\right.
\label{eq_harmonic_average}
\end{equation}
which are approximated in practice with Discrete Fourier Transforms (DFTs, see section \ref{sec_results_testbench} and Appendix \ref{appendix_identification}).
Also, as the forcing frequency $k\omega_u$ is chosen outside the frequency support of $y(t)$ we have:
\begin{equation}
\widehat{\delta y}_{\mathbf{\vect{\Phi}}}(k\omega_u) = \widehat{y}_{\mathbf{\vect{\Phi}}}(k\omega_u).
\end{equation}
This is particularly important since we wish to identify the transfer between the input and the linear part of the output, that is not easily measurable in practice.
Then, a frequency response depending on the phase $\mathbf{\vect{\Phi}}$ of the input, may be computed as a ratio of Fourier coefficients at forced frequencies:
\begin{equation}\label{eq_fourier_ratio}
H_{\mathbf{\vect{\Phi}}}(\jwu)  = \frac{\widehat{\delta y}_{\mathbf{\vect{\Phi}}}(k\omega_u)}{\widehat{u}_{\mathbf{\vect{\Phi}}}(k\omega_u) }.
\end{equation}
Now, using the expression of $y_{\mathbf{\vect{\Phi}}}(t)$ deduced from \cite{leclercq2023}, it can be shown that:
\begin{equation}
\Ep_{\mathbf{\vect{\Phi}}}(H_{\mathbf{\vect{\Phi}}}(\jwu))  = H_0(\jwu).
\end{equation}
Therefore, if the experiment is repeated over $M$ realizations of $u_{\mathbf{\vect{\Phi}}}(t)$ (i.e. by sampling $\mathbf{\vect{\Phi}}$), then the sample mean $\bar{H}_{\mathbf{\vect{\Phi}}}$ of $H_{\mathbf{\vect{\Phi}}}$ approximates $H_0$ with variance $\Var(\bar{H}_{\mathbf{\vect{\Phi}}}(\jwu)) = \frac{1}{M}\Var(H_{\mathbf{\vect{\Phi}}}(\jwu))$. 
It is notable that the ensemble average is done here on the multisine phase $\mathbf{\vect{\Phi}}$, and not the phase $\phi$ on the limit cycle where the signal $u_{\mathbf{\vect{\Phi}}}(t)$ is injected (which was done in \cite{leclercq2023}). 
Here, the phase on the limit cycle $\phi$ is assumed constant. 
In an experiment where $\phi$ cannot be chosen, the ensemble average would rather be performed on $\phi$ only, maintaining $\mathbf{\vect{\Phi}}$ constant (i.e. injecting the same input signal at different instants on the limit cycle). As such, it would be possible to obtain a sample average of $\Ep_{\phi}(H_{\phi}(\jwu))  = H_0(\jwu)$.

\subsubsection{System identification}
Now that $H_0(\jw)$ has been sampled on a grid of $\omega$, we wish to find a low-order state-space representation of the transfer function $G(s)$ approximating the unknown $H_0(s)$, accessible to control techniques. 
This task is performed with a subspace-based frequency identification method, but could be performed with any other frequency identification method, e.g. the Eigensystem Realization Algorithm in Frequency-Domain (ERA-FD, \cite{ERAFD}), or vector-fitting \citep{vectorfitting,vectorfitting2}. 
Subspace methods form a class of blackbox linear identification methods, that do not rely on nonlinear optimization as most iterative model-fitting methods do. 
Here, the Frequency Observability Range Space Extraction (FORSE, \cite{forse}) estimates a discrete-time state-space model with order fixed a priori, in two distinct steps. Matrices $\mat{A}$ and $\mat{C}$ are built directly from a SVD of a Hankel matrix built from the frequency response. Matrices $\mat{B}$ and $\mat{D}$ are then found by solving a linear least squares problem. More details can be found in \cite{forse} or in Appendix \ref{appendix_forse} for a SISO version. Additional stability constraints may be enforced with Linear Matrix Inequalities (LMIs, \cite{fabricelmi}), as the transfers $H_0(s), G(s)$ are expected to exhibit some marginally-stable poles in this context.

For the sake of rendering the procedure as automatic as possible, the order of the model identified at each iteration, denoted $n_G$, is fixed. The choice of $n_G$ is discussed in \S \ref{sec_results_testbench}.

%%%%%%%%%%%%%%%%%%%%%%%%%%%%%%%%%%%%%%%%
\subsection{Control of the flow using the mean transfer function}\label{sec_control}
\subsubsection{Rationale}
After we have determined an LTI model $G(s)$ of the fluid around its attractor, we wish to control it in order to reduce the self-sustained oscillations. 
Among the classic control methods such as pole placement, LQG, $\Hinf$ techniques (e.g. mixed-sensitivity, loop-shaping, structured $\Hinf$,...), and MPC, we choose the LQG framework for synthesis. 
It combines several advantages, such as being easy to automate, 
having predictable controller gain to some extent, 
and producing a controller with relatively low complexity.

\subsubsection{Principle of observed-state feedback}\label{sec_lqg}
Linear Quadratic Gaussian control is very popular in flow control (see e.g. \cite{schmid2016,barbagallo2009,bewleylinear,carini2015,brunton2015} and references therein) due to its ease-of-use, but the principle is recalled for SISO plants. Consider a plant $G = \statespace{A}{B}{C}{D}$ with state vector $\vect{x}$, control input $u$ and measurement output $y$.

\paragraph{Linear Quadratic Regulator (LQR).}
We wish to solve an \emph{optimal control} problem with full-state information, i.e. find the control input $u(t)$ that minimizes a performance criterion defined as $J = \int_0^\infty [ \vect{x}(t)^T\mat{Q} \vect{x}(t) + {R}{u^2}(t) ] \, \d t$ with parameters $R>0, \mat{Q} \succeq 0$ (positive semi-definite matrix). The solution is a state feedback law ${u}(t)=\mat{K}\vect{x}(t)$, where $\mat{K}$ is a matrix of gains computed as $\mat{K}=-{R}^{-1}\mat{B}^T\mat{P}$, with $\mat{P}$ the solution of the Riccati equation: $\mat{A}^T\mat{P} + \mat{P}\mat{A} - \mat{P}\mat{B}{R}^{-1}\mat{B}^T\mat{P}+\mat{Q}=0$. 

\paragraph{Construction of an observer.}
Here, the state $\vect{x}(t)$ of the reduced-order plant $G$ cannot be accessed directly, and only the output $y(t)=\mat{C}\vect{x}(t)$ is available for feedback control. We construct an estimate $\hat{\vect{x}}(t)$ of the state of the process based on past measurements and inputs, and use it in an observed-state feedback law: $u(t)=\mat{K}\hat{\vect{x}}(t)$.
The dynamics for the estimated state $\hat{\vect{x}}$ reproduce the plant dynamics, with a corrective forcing $\hat y - y$ weighted by a gain $\mat{L}$ to be defined, accounting for measurements in real-time:
\begin{equation}
\left\{
\begin{aligned}
& \dot{\hat{\vect{x}}} = \mat{A} \hat{\vect{x}} + \mat{B}u + \mat{L}^T(\hat y - y),  \\
& \hat y =  \mat{C} \hat{\vect{x}}.
\end{aligned}
\right.
\end{equation}
It can be shown that solving for $\mat{L}$ is dual to the previous problem, with $\mat{A}\leftarrow \mat{A}^T, \mat{B}\leftarrow \mat{C}^T$ and new parameters $\mat{W} \succeq 0, {V} > 0$ that are covariances of additive white noise on the state $\vect{x}(t)$ and measurement $y(t)$. 

\paragraph{Linear Quadratic Gaussian Regulator.}
The final controller is formed with an observer gain $\mat{L}$ and a state-feedback gain $\mat{K}$, making it dynamic and expressed as follows:
\begin{equation}
K_{LQG}(s) =  \statespace{A+BK + L^TC}{ -L^T}{K}{0}.
\end{equation}

\paragraph{Remarks on the LQG controller.}
\begin{enumerate}
\item~ The state-feedback gain $\mat{K}$ and the observer gain $\mat{L}$ are found independently
Additionally, each problem may be normalized. For the state-feedback, the state weighting is chosen as $\mat{Q}=\mat{I}_n$ and the problem is only parametrized by the value $R > 0$ penalizing the control input. Symmetrically, for the estimation problem, we choose $\mat{W} = \mat{I}_n$ and the problem is parametrized by $V > 0$. This approach is more conservative because states are weighted equally, but allows for parametrizing the LQG problem easily with only two positive scalars $R, V > 0$.
                        
\item~ The matrix weights may be tuned to enforce specific behaviors for the solution controller (which was already underlined in \cite{sipp2016}). For the state-feedback problem, one can prioritize small control inputs ($R\to \infty$, low gain $\mat{K}$) or reactive control ($R\to 0$, large gain $\mat{K}$). Symmetrically, for the estimation problem, the choice is made between slow estimation ($V\to \infty$, low gain $\mat{L}$)  or fast estimation ($V \to 0$, large gain $\mat{L}$). 

\item~ While LQR controllers exhibit inherent robustness (in gain and phase margins) given diagonal $\mat{Q}, \mat{R}$ \citep{lqr_robustness}, these guarantees generally do not hold for LQG and stability margins may be arbitrarily small \citep{doyle1978guaranteed} but are checked a posteriori.
\end{enumerate}

%%%%%%%%%%%%%%%%%%%%%%%%%%%%%%%%%%
\subsection{Controller stacking, balanced truncation and state initialization}\label{sec_reduction}
\subsubsection{Rationale}
At the beginning of iteration $i+1$ of the procedure, a new controller $K_{i}^+$ is synthesized and coupled to the flow, that is already in closed-loop with the control law $K_{i}$. The new controller operating in the loop would be:
\begin{equation} 
K_{i+1} = K_{i} + K_{i}^{+}.
\end{equation}
In the general case, while the newly-designed controller $K_{i}^+$ has manageable order ($\ord{K_i^+}=\ord{G_i}=n_G$), the total controller $K_{i+1}$ has order $\ord{K_{i+1}} = \ord{K_i}+\ord{K_i^+} =(i-1) n_G + n_G$ that is increasing linearly with iterations.

In order to reduce the order of the controller, we resort, at each iteration, to balanced truncation of the controller operating in the loop.
In other words, instead of using the full controller $K_{i+1}= K_{i} + K_{i}^{+}$ in the loop, we use a reduced-order version $\tilde{K}_{i+1}$. Repeating the operation at each iteration leads to:
\begin{equation}
\tilde{K}_{i+1} = \BT(\tilde{K}_{i}+ K_i^+),
\label{eq_ki_redux}
\end{equation}
 where the operation $\BT$ refers to the balanced truncation described below, enabling order reduction: $\ord{\left[\BT(K)\right]} \leq \ord{K}$. This operation makes state initialization of the new controller more challenging, which is tackled in the following as well.

\subsubsection{Balanced truncation} \label{sec_bt}
Balanced truncation was already introduced in previous flow control articles \citep{rowley2005model,bewleylinear} with the intent to reduce the order of flow models. As the traditional balanced truncation algorithm introduced in \cite{moore1981} is not applicable to high-dimensional models of order $O(10^5)$, it has led to the development of various approximate techniques. However, the objective is different here: order reduction is performed on the controller which initially has moderate order $O(10)$, enabling direct balanced truncation methods (see \cite{balrealsurvey} for an in-depth survey).

More specifically, given a controller $K$
of order $\ord{K} = n_K$, we wish to find a reduced-order controller $\tilde{K}=\BT(K)$
of order $\ord{\tilde K} = n_r<n_K$ such that the controllers $K, \tilde K$ have similar behavior, quantified as the $\Hinf$ norm of the difference.
This is done by first computing a balanced realization of $K$ \citep{moore1981,balreal1}, then truncating the balanced modes of $K$ with the lowest controllability and observability, quantified by its Hankel Singular Values (HSV), denoted $\sigma_j, j\in \{1,\dotsc, n_K\}$ \citep{pernebo1982}. For a truncation to order $n_r$, the reduction error is bounded explicitly \citep{enns1984}: 
\begin{equation}
\norm{K(s) - \tilde{K}(s)}_\Hinf  \leq 2 \sum_{j=n_r+1}^{n_K} \sigma_j.
\label{eq_bt_bound}
\end{equation}
 If $K$ has unstable modes (which cannot be prevented in the LQG framework), a possibility is to separate $K$ into unstable and a stable contributions $K=K_u + K_s$, and perform the reduction on $K_s$ only \citep{zhou1999}.

\subsubsection{Controller switching} \label{sec_controller_switching}
When adding a new controller to the flow, careful state initialization is needed in order to reduce transients in the control input (e.g. so as to not saturate the actuator in a real-life setup or to not trigger nonlinear effects).
Many techniques from the control literature address smooth switching, either with \emph{bumpless} methods (usually requiring a precise full plant model, \cite{bumpless_zaccarian}), by focusing on fast controller transients (\cite{bumpless_safonov}, see also references therein for a brief overview), or by using past data for initializing the controller \citep{paxmanphd}. 
Although these techniques are very attractive, they did not seem to provide consistent performance on our study case. The explanation might come from the fact that the flow model is very crude by definition: nonlinearity is neglected, and the time-dependent variations of the input-output transfers are averaged in the mean transfer function.
We present below a basic two-step method that proved to be consistent and efficient in this study. 

The controller initialization and switching is done in two steps: first, basic state initialization for the full-order controller in order to reach a new dynamical equilibrium with moderate transient; then, state initialization of the reduced controller based on the past transient signal, ensuring seamless transition between the high-order and low-order controllers. 
For simplicity of notations, we redefine time with respect to the current iteration: $t=0$ is the time instant where the full-order controller $K_{i+1}$ is inserted, and $t=T_{sw}$ is the instant where $K_{i+1}$ and its reduced-order counterpart $\tilde K_{i+1} = \BT(K_{i+1})$ are exchanged.

\paragraph{First step: full-order controller state initialization.}
Without controller order reduction, the new controller $K_i^+$ is added to the flow with internal state $\vect{x}_{K_i^+}^0={0}$. The current internal state $\vect{x}_{\tilde{K}_i}$ of the closed-loop controller $\tilde{K}_i$ is untouched, so that the stacked controller $K_{i+1}$ has initial internal state :
\begin{equation}
\vect{x}_{K_{i+1}}^0= \begin{bmatrix} \vect{x}_{\tilde{K}_i} \\  0 \end{bmatrix},
\label{eq_xk0_full}
\end{equation}
as in \cite{leclercq2019}. 
Initialized as such, the controller in closed-loop generally produces a control input $u(t)$ and a corresponding measurement signal $y(t)$ with moderate transient amplitude, thanks to the choice of LQG weightings detailed in \S\ref{sec_lqg_weightings}. Once the transient term related to controller state initialization has decayed, after a duration $T_{sw}$, we perform a seamless switch to the reduced-order controller.

\paragraph{Second step: reduced-order controller state initialization and switch.}
Just as the previous step, this step is still a controller initialization problem: how should the state of the reduced-order controller $\vect{x}_{\tilde{K}_{i+1}}$ be set in order to produce small control transient when exchanging controllers in the loop at time $t=T_{sw}$? 
To solve this, we exploit the fact that both controllers produce close outputs $u(t), \tilde u(t)$ under the same history of input signal $y(\tau), 0 \leq \tau \leq t$. Indeed, by definition of the $\Hinf$ norm, for $\norm{y(t)}_{2} > 0$: 
\begin{equation}
\frac{\norm{u(t) - \tilde u(t)}_{2}}{\norm{y(t)}_{2}} \leq \norm{K_{i+1}(s) - \tilde{K}_{i+1}(s)}_\Hinf,
\end{equation}
where $\norm{K_{i+1}(s) - \tilde{K}_{i+1}(s)}_\Hinf$ is bounded from \eqref{eq_bt_bound} and expected to be small due to the truncation of balanced modes with small HSVs. As a result, $u$ and $\tilde u$ may be expected to be close under the same input $y$. 
Then, if we denote the full-order controller $K_{i+1}=\statespace{A_K}{B_K}{C_K}{0}$, we may write its output as: 
\begin{equation}
u(t) = \mat{C_K} e^{\mat{A_K}t} \vect{x}^0_{K_{i+1}}  + \int_0^t \mat{C_K}e^{\mat{A_K}(t-\tau)}\mat{B_K} y(\tau) \, \d \tau,
\end{equation}
where $y$, arising from the closed-loop, can be considered an external signal to the controller. 
If the reduced-order controller $\tilde K_{i+1}=\statespace{\tilde{A}_K}{\tilde{B}_K}{\tilde{C}_K}{0}$ were fed the same signal $y$, but only $u$ would be fed back to the flow (i.e., not $\tilde{u})$, then its output signal would be equivalently:
\begin{equation}
\tilde u(t) = \mat{\tilde{C}_K} e^{\mat{\tilde{A}_K}t} \vect{x}^0_{\tilde K_{i+1}}  + \int_0^t \mat{\tilde{C}_K} e^{\mat{\tilde{A}_K}(t-\tau)}\mat{\tilde{B}_K} y(\tau) \, \d \tau.
\label{eq_utilde_tw}
\end{equation}

Assuming $T_{sw}$ is chosen such that the initial condition contribution becomes negligible, then the state of the reduced-order controller producing the output \eqref{eq_utilde_tw}, defined as $\tilde{u}(T_{sw}) = \mat{\tilde{C}_K} \vect{x}_{\tilde{K}_{i+1}}\!(T_{sw})$, is equal to:
\begin{equation} 
\vect{x}_{\tilde{K}_{i+1}}(T_{sw}) = \int_{0}^{T_{sw}} e^{\mat{\tilde{A}_K}(T_{sw}-\tau)}\mat{\tilde{B}_K} y(\tau) \, \d \tau.
\label{eq_xk_switch}
\end{equation}
Therefore, the two controllers are switched at time $t=T_{sw}$, by setting the state of the reduced-order controller as per \eqref{eq_xk_switch}. More importantly, the closed-loop behavior of the measurement signal $y$ for $\tau \geq T_{sw}$ is expected to remain similar since the two controllers are designed to show similar input-output behavior. In this study, we have set $T_{sw}=50$ which is usually larger than the characteristic time scales of the controllers (computed from the eigenvalues of $\mat{A_K}, \mat{\tilde{A}_K}$). 

\begin{figure}
\centering
\includegraphics[width=0.8\textwidth]{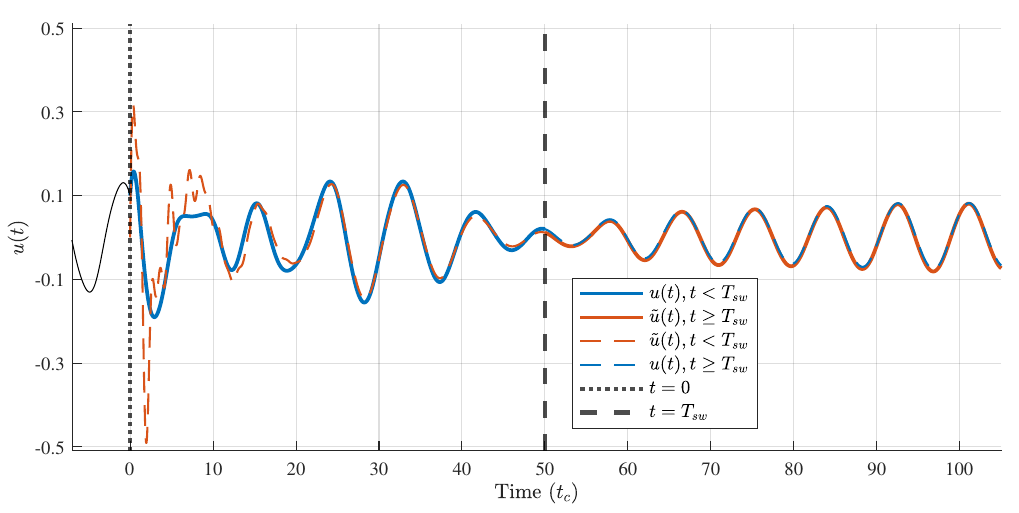}
\caption{Controller initialization and switching procedure. The control signal used is first $u(t), t<T_{sw}$ (solid blue) generated by the full-order controller, initialized as \eqref{eq_xk0_full}; then, it is switched to $\tilde{u}(t), t\geq T_{sw}$ (solid red) generated by the reduced-order controller, initialized as \eqref{eq_xk_switch}. The dashed signals need not be computed in practice, but are represented nonetheless.}
\label{fig_u_switch}
\end{figure}

\paragraph{Illustration.} The two-step switching process is illustrated in figure \ref{fig_u_switch}, with data from \S\ref{sec_results_control}. For the first step, the signal $u(t)$ (solid blue) generated by the full-order controller $K_{i+1}$ is used in closed-loop for $t<T_{sw}$, with initial state $\vect{x}_{K_{i+1}}^0$ as per \eqref{eq_xk0_full}. For the second step, for $t\geq T_{sw}$ the signal $\tilde{u}(t)$ (solid red) generated by the reduced-order controller $\tilde{K}_{i+1}$ is used in place of $u(t)$, by setting its internal state as per $\eqref{eq_xk_switch}$. 
For representation purposes, we compute and represent the signals $\tilde{u}(t), t<T_{sw}$ and $u(t), t\geq T_{sw}$, as dashed red and blue lines, respectively. In practice, they need not be computed. Figure \ref{fig_u_switch} confirms the benefit of this switching procedure: the transient regime is moderate and the transition from the full-order to the reduced-order controllers is almost seamless.

%%%%%%%%%%%%%%%%%%%%%%%%%%%%%
\section{Results: driving the flow from the limit cycle to the base flow}\label{sec_results}

\subsection{Unforced flow: a test-bed for choosing parameters} \label{sec_unforced_flow}
The first iteration for the identification-control procedure starts at time $t_1=500$, when the unforced flow is fully-developed, as no controller is in the loop yet. The unforced fully-developed flow at $t_1=500$ is exploited in order to evaluate the sensitivity of the method to its parameters.

\subsubsection{Unstable equilibrium and unforced flow}
\begin{figure}
\centering
\includegraphics[width=0.7\textwidth]{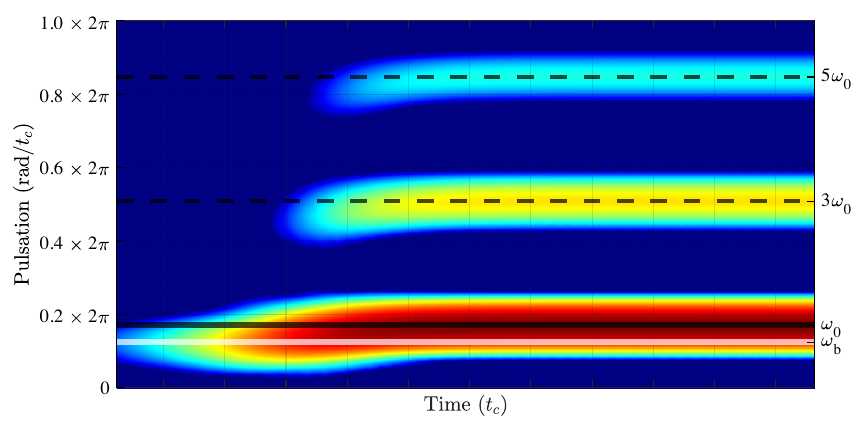}
\caption{Spectrogram of cross-stream velocity probe at $x_1=3, x_2=0$ (used for feedback), flow trajectory from unstable equilibrium to natural stable limit cycle. The pulsation continuously shifts from $\omega_b$ to $\omega_0$, and higher-order odd harmonics gradually appear.}
\label{fig_spectro_div}
\end{figure}

To compute the fully-developed regime of self-sustained oscillations, the flow initially on its unstable equilibrium, is perturbed infinitesimally to leave the equilibrium. 
The fundamental oscillation frequency of the flow continuously shifts from the frequency of the unstable pole of the flow linearized around the equilibrium $\Gb(s)$: $\omega_b = 0.779  \, \rad/t_c$, to the frequency of the limit cycle $\omega_0 = 1.062  \, \rad/t_c$ on the attractor (see spectrogram in figure \ref{fig_spectro_div}). Also, it is notable that almost no high-order harmonics are visible at small time instants near the equilibrium, which is a sign of the complex exponential divergence; while the several higher-order harmonics appear due to the nonlinearity at greater time instants. This phenomenon is also more visible in the far wake. Note that by half-wave symmetry of the unforced signal $y(t)$ (i.e. $y(t) = -y(t+\frac{T}{2})$), only odd harmonics appear in its frequency representation. This symmetry of the cross-stream velocity component on the symmetry axis of the geometry is justified in \cite{barkley2000bifurcation} by the alternance of vortex shedding from the upper and lower mixing layers over a period.

\subsubsection{Identification: multisine design and reduced-order model} \label{sec_results_testbench}
\paragraph{Design of input signal.}
The first step of the iterative process is identifying the mean transfer function of the flow thanks to multisine excitations, as per \ref{sec_identification}.
The fundamental pulsation $\omega_u$, that enables gathering the frequency response at pulsations $k\omega_u$, is chosen as $\omega_u=2\pi \cdot 10^{-2}$, providing fine enough sampling of the frequency response, especially near resonant modes. Consequently, the input signal is periodic with period $T_u = 100$. The $N=5000$ frequencies included in the input signal are such that the highest frequency is $N \omega_u = 2\pi \times \frac{1}{4} f_s$ (where $f_s= 200$ is the sampling frequency), in order to cut high-frequency harmonics ($f > \frac{1}{4} f_s$) that are weakly amplified by the flow, therefore not useful for identification nor control.

For the first iteration, we choose the amplitude of the input $u_{\mathbf{\vect{\Phi}}}(t)$ as $A=10^{-3}$ to guarantee small perturbation, i.e. $\norm{\delta y_{\mathbf{\vect{\Phi}}}(t)}_{\infty} \ll 1$. 
However, the closer the system is to the equilibrium (i.e. at higher iterations of the procedure), the more $A$ is reduced because the relative impact of forcing increases. Note that this would need to be taken into account in an experiment, where the output noise may be important, so there is a balance to strike between small input $u_{\mathbf{\vect{\Phi}}}(t)$ and convenient signal-to-noise ratio. 

The mean frequency response $\bar{H}_{\mathbf{\vect{\Phi}}}(\jwu)$ is estimated by averaging $M=4$ realizations of $H_{\mathbf{\vect{\Phi}}}(\jwu)$. The choice of the value for $M$ is justified in Appendix \ref{appendix_identification}, where we also provide some details on the estimation in practice.
For each realization, the forced system is simulated on a duration $(P_{tr} + P)T_u$, with $P_{tr}=P=4$. The first portion of the input and output signals of duration $P_{tr} T_u$ is discarded for containing the contribution of damped Floquet modes in the flow response \citep{leclercq2023}. The second portion of duration $PT_u$ is utilized for estimating  $H_{\mathbf{\vect{\Phi}}}(\jwu)  = \frac{\widehat{\delta y}_{\mathbf{\vect{\Phi}}}(k\omega_u)}{\widehat{u}_{\mathbf{\vect{\Phi}}}(k\omega_u) }$ \eqref{eq_fourier_ratio}.

\begin{figure}
\centering
\includegraphics[width=0.7\textwidth]{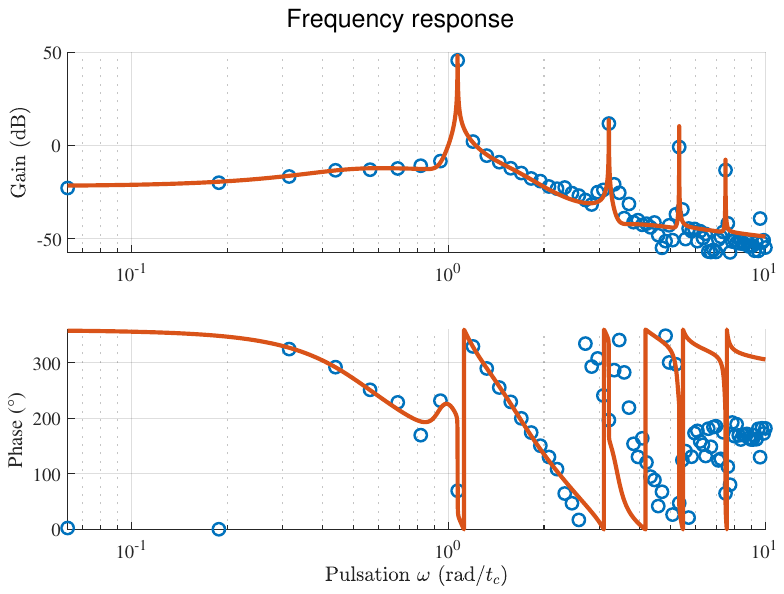}
\caption{Mean frequency response (blue circles) and identified ROM $G_0$ (solid red line) at the first iteration.}
\label{fig_identification_G0}
\end{figure}

\paragraph{Identified mean transfer function.}
The mean frequency response and the associated identified low-order model $G_0$ at the first iteration are represented in figure \ref{fig_identification_G0} (as blue circles for the frequency response, and as a solid red line for $G_0$). 
The ROM shows good fit with respect to the frequency data at low frequency, and manages to recover the resonant poles as well as the linear slope of the phase in the pulsation range $\omega \in [1.1, 3] \, \rad/t_c$. 
For $\omega \geq 3$, the ROM does not match the frequency response as closely, which is especially visible on the phase plot. It is explained by appendix \ref{appendix_identification}, suggesting the need for more realizations of $H_{\mathbf{\vect{\Phi}}}(\jwu)$ to converge the sample mean at higher frequencies.

\paragraph{Order of the ROM.} In order to avoid overfitting the frequency response (which is not always fully converged due to a lack of data), the order is chosen as low as $n_G=8$ at each iteration.
Identifying greater orders tends to introduce spurious modes into the ROM, and keeping a low order proves to still capture the main features of the flow response (resonant modes and unstable zeros). The mean transfer function is expected to have poles on the imaginary axis \citep{leclercq2023}, so the identification algorithm is tuned to enforce $\Re(p_0)\leq 0$ for all poles $p_0 \in \mathbb{C}$ thanks to Linear Matrix Inequalities constraints \citep{fabricelmi}. Numerically, the poles are usually found in the strictly stable half-plane ($\Re(p_0) < 0$), but marginally-unstable poles would not pose an issue for control in this case.

\subsubsection{Control design: choice of LQG weights}\label{sec_lqg_weightings}
By construction, the mean transfer function ROM $G$ is valid for small control input (and weakly unsteady flow in the sense that the time-varying nature of the flow may be safely neglected from an input-output viewpoint), which is an incentive to design a controller with small gain. With LQG synthesis, it corresponds to the limit $R\to \infty$ (high penalization of control input, slow controller) and $V\to \infty$ (noisy measurement, slow observer).
However, for infinitesimal control input, while the LTI approximation is valid, it is likely that the control produces imperceptible change in the flow. On the contrary, as the gain increases, modification of the flow should be more and more discernible, until the nonlinear phenomena becoming more present and the assumption of linearity fails.

\begin{figure}
\centering
\begin{subfigure}{.48\textwidth}
\centering
\includegraphics[height=5.5cm]{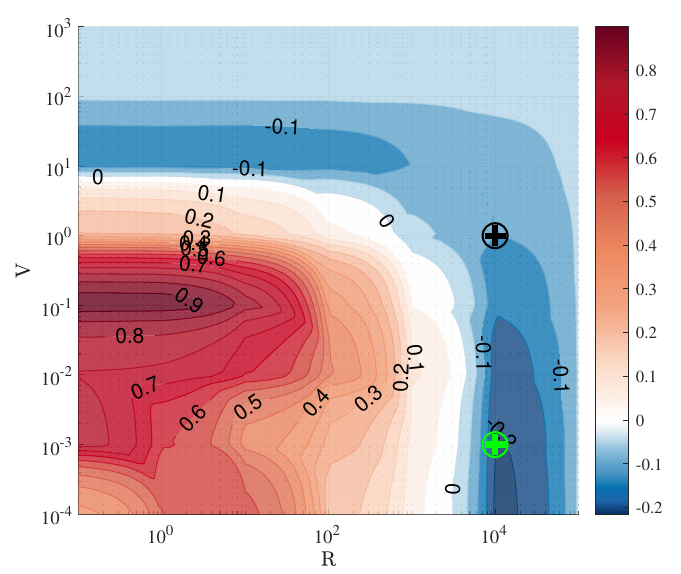}
\caption{PKE criterion $\delta\Ek_1$.}
\label{fig_map_lqg_1}
\end{subfigure}
\begin{subfigure}{.48\textwidth}
\centering
\includegraphics[height=5.5cm]{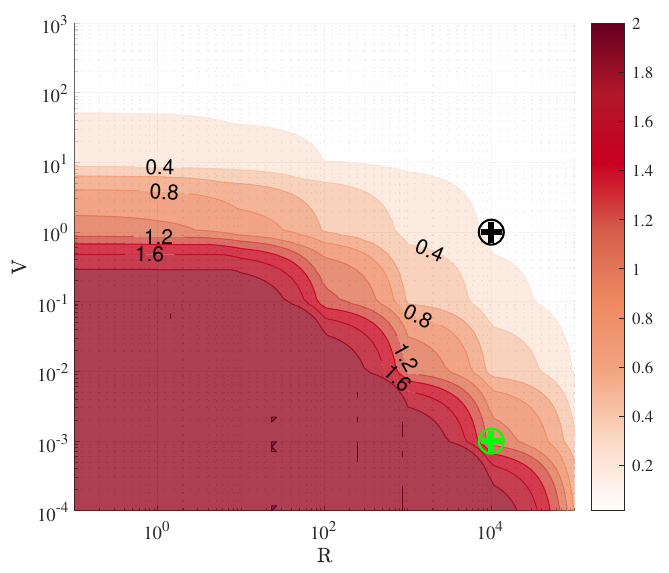}
\caption{Maximum amplitude of input $\max_t |u(t)|$.}
\label{fig_map_lqg_2}
\end{subfigure}
\caption{Meshing of parameters $R, V$ for LQG synthesis at the first iteration. A controller is deemed satisfactory when it achieves PKE reduction (figure \ref{fig_map_lqg_1}) and moderate control input transient (figure \ref{fig_map_lqg_2}). In this study, the chosen parameters are indicated as crosses (green for the first iteration, black for subsequent iterations).}
\label{fig_map_lqg}
\end{figure}

In order to choose the more appropriate weights for the LQG controller, we coarsely mesh $R, V$, generate the associated controllers $K(R, V)=LQG(G, R, V)$ and plug them in the fully-developed flow. The flow in feedback with the control law $K(R,V)$ reaches a new dynamical equilibrium after some transient regime. 
A controller is deemed satisfactory if it achieves both of the following:
\begin{itemize} 
\item~ Average perturbation kinetic energy (defined in \S \ref{sec_notations_flow}) on the new dynamical equilibrium in closed loop $\Ek_1(R,V)$ lower than the natural limit cycle PKE $\Ek_0$. This criterion is conveyed by $\delta\Ek_1(R,V) = \frac{\Ek_1(R,V)}{\Ek_0} - 1$. A map of $\delta\Ek_1(R,V)$ is shown in figure \ref{fig_map_lqg_1}. Desirable control laws should yield $\delta\Ek_1(R,V) < 0$.

\item~ Moderate control input $u(t)$ in order to avoid potential actuator saturation, conveyed by the quantity
$\max_t |u(t)|$,
in figure \ref{fig_map_lqg_2}.
\end{itemize}

The low-gain control region, with expensive control $R\to \infty$ and slow estimation $V\to \infty$, corresponds to the upper-right corners in figures \ref{fig_map_lqg_1} and \ref{fig_map_lqg_2}. On the contrary, the high-gain control region, with cheap control $R\to 0$ and fast estimation $V\to 0$, corresponds to the lower-left corners of the same figures.
As expected, for low-gain controllers, almost no modification in the PKE is observed (upper-right corner in figure \ref{fig_map_lqg_1}). When increasing the gain (either decreasing $R$, or $V$, or both), energy reduction becomes increasingly more perceptible, until the performance ultimately degrades, most probably due to nonlinearity, corresponding to the limit of validity of the mean transfer function (lower-left corner in figure \ref{fig_map_lqg}). 
It appears clearly that the low-gain limit (small PKE reduction) is achieved with either $R\to \infty$ (right-side boundary of \ref{fig_map_lqg_1}) corresponding to low control gains $\mat{K}$, or $V\to \infty$ (upper boundary of \ref{fig_map_lqg_1}) corresponding to low estimation gains $\mat{L}$. This limit may be thought as a safe zone for choosing parameters (since a controller would likely not disturb the flow perceptibly) but not guaranteeing sufficient PKE reduction.

For the first iteration, assuming that the maximum value of the control input that can be implemented in simulation  (corresponding to an actuator saturation in an experiment) is around $\max_t |u(t)|\approx 2$, values of $R, V$ are chosen as $R=10^{4}, V=10^{-3}$ (green cross) that offer PKE reduction of $\delta\Ek_1 \approx -20\%$ and moderate control input, without exceeding the arbitrary saturation. For subsequent iterations, parameters are chosen more conservatively because energy maps can be expected to vary slightly: $R = 10^{4}, V=10^{0}$ (black cross).

\subsubsection{Controller reduction: choice of the balanced truncation threshold}
With $K$ in balanced form with Hankel singular values (HSV) sorted by decreasing magnitude, states with HSV $\sigma_j$ such that $\frac{\sigma_j}{\sigma_1}  < g_\sigma$ (with chosen threshold $g_\sigma$) are truncated. 
In our study case, $g_\sigma=10^{-3}$  provides a good trade-off between maintaining performance and sufficient order reduction, usually such that $n_r \leq 25$. Note that a fixed maximum number of states could be forced, with the risk of truncating important dynamics in the reduced-order controller.

\subsection{Main result: data-based convergence to equilibrium with piecewise-LTI controller}
The main result of this study is the stabilization of the cylinder from its natural limit cycle to its unstable equilibrium, using only input-output data and LTI controllers, in a way that could be implemented in an experiment.
In the results presented here, the stabilization took $8$ iterations from the limit cycle in order to attain the equilibrium, stabilized by a closed-loop LTI controller.

%%%%%%%%%%%%%%%%%%%%%%%%%%%%%%%%%
\subsubsection{Iterative convergence to the base flow}
The convergence to the base flow is shown through the perturbation kinetic energy (PKE), defined 
in \S \ref{sec_notations_flow}. Although this quantity is not available outside the simulation realm, it is used as a posterior criterion of convergence to the base flow. 
\begin{figure}
\centering
\includegraphics[width=0.7\textwidth]{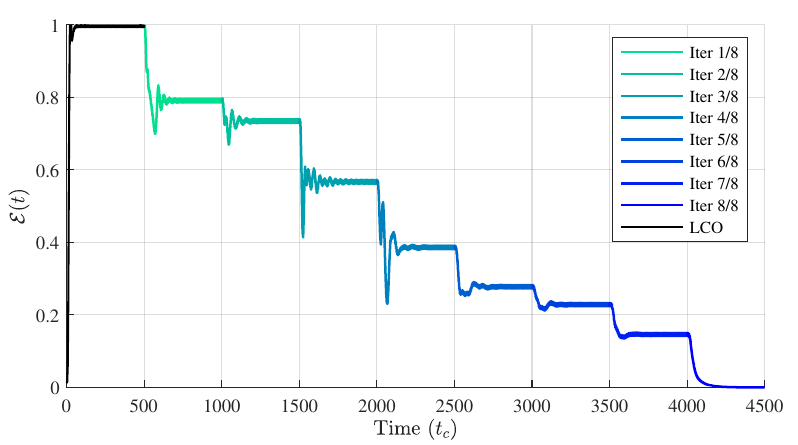}
\caption{PKE $\Ek(t)$ throughout iterations.}
\label{fig_dE}
\end{figure}
At each iteration, it is observed in figure \ref{fig_dE} that the PKE $\Ek$ decreases to reach a new dynamical equilibrium with time-averaged value $\langle \Ek(t) \rangle_t$ (normalized by the PKE on the attractor $\Ek_0$). At the last iteration starting at $t=4000$, the flow is attracted to the base flow. We have $\lim_{t\to \infty} \Ek(t) = 0$, and the convergence is exponential: the system is stabilized around the base flow in closed-loop in a linear sense. It is notable that the approach is based solely on input-output data, so the PKE would not be ensured to decrease at each iteration (and other solutions to the problem showed that convergence can be achieved without monotonic decrease of the average PKE at each iteration).

%%%%%%%%%%%%%%%%%%%%%%%%%%%%%%%%%
\subsubsection{Probing the wake}
\begin{figure}
\centering
\includegraphics[width=0.8\textwidth]{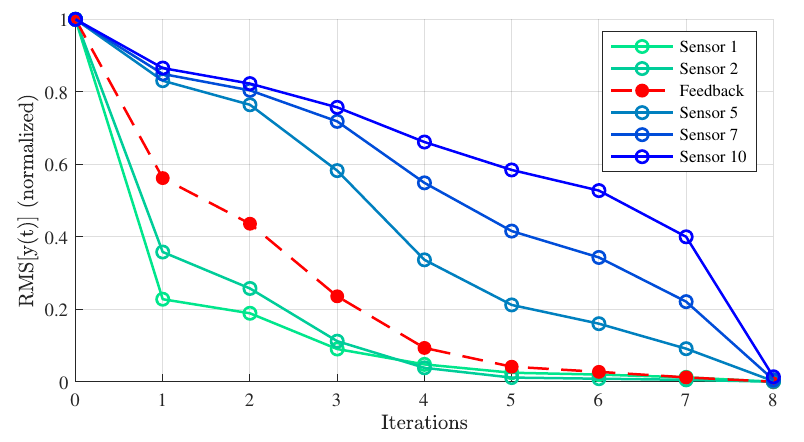} 
\caption{Normalized RMS of measured signals (performance sensors) depending on position in the wake, throughout iterations.}
\label{fig_yrms_sensors}
\end{figure}

The full stabilization of the flow is confirmed by inserting probes in the flow (not used for feedback), on the symmetry axis $x_2=0$, at $x_1=1, 2$ (upstream of the feedback sensor) and $x_1 = 5, 7, 10$ (downstream of the feedback sensor). Additionally, it enables tracing profiles of convergence of the cylinder wake: in figure \ref{fig_yrms_sensors}, we compute the RMS value of the signal at each probe location, normalized by the RMS of the same probe in the unforced flow. Closer sensor signals ($x_1 \leq 3$) have normalized RMS decay very rapidly and reach $10\%$ as soon as iteration $4$, while for downstream probes, the signal reduces very mildly at first. It corresponds to the incremental lengthening of the recirculation bubble behind the cylinder, so that closer sensors quickly lie in the mean recirculation bubble, while downstream sensors still probe a strongly oscillating wake.
It would be interesting to apply the procedure with various feedback sensor positions, in order to identify the zones where  the feedback sensor is the most effective in a nonlinear setting.

\begin{figure}
\centering
\begin{subfigure}{.3\textwidth}
  \centering
  \includegraphics[width=\textwidth, trim={4cm 3cm 2cm 3cm}, clip]{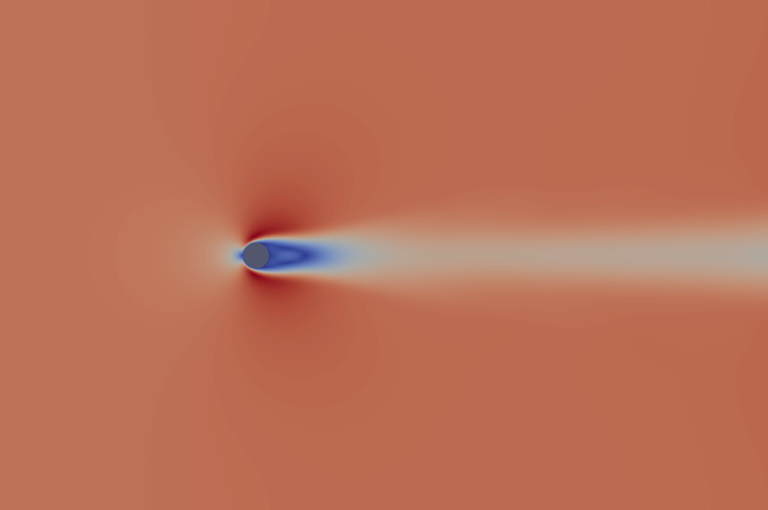} 
\end{subfigure}
\begin{subfigure}{.3\textwidth}
  \centering
  \includegraphics[width=\textwidth, trim={4cm 3cm 2cm 3cm}, clip]{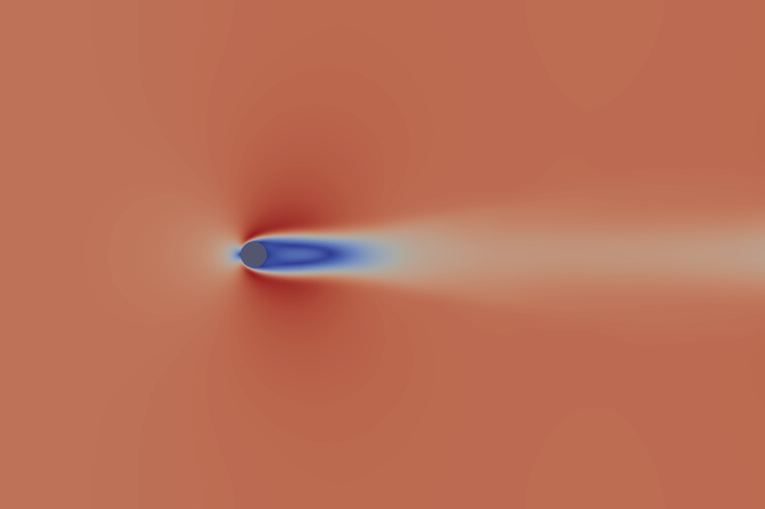} 
\end{subfigure}
\begin{subfigure}{.3\textwidth}
  \centering
  \includegraphics[width=\textwidth, trim={4cm 3cm 2cm 3cm}, clip]{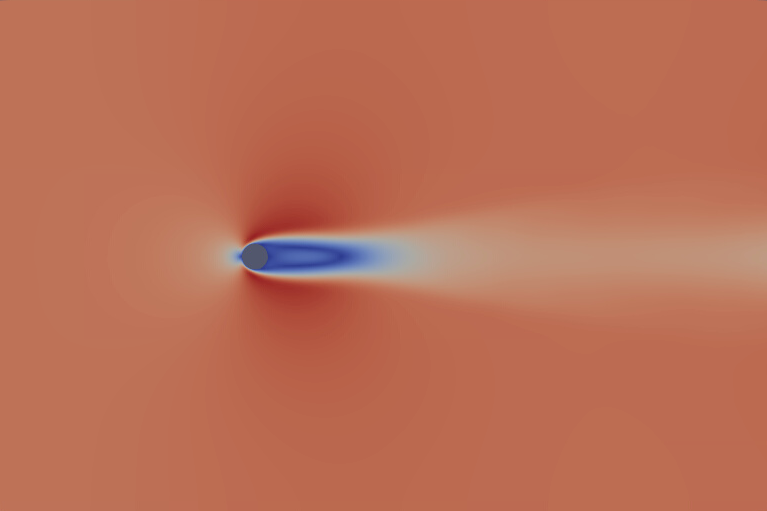} 
\end{subfigure}
\begin{subfigure}{.3\textwidth}
  \centering
  \includegraphics[width=\textwidth, trim={4cm 3cm 2cm 3cm}, clip]{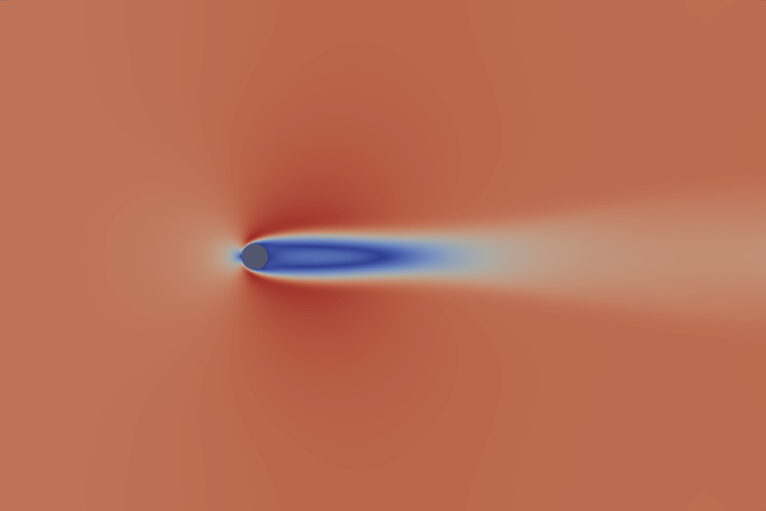} 
\end{subfigure}
\begin{subfigure}{.3\textwidth}
  \centering
  \includegraphics[width=\textwidth, trim={4cm 3cm 2cm 3cm}, clip]{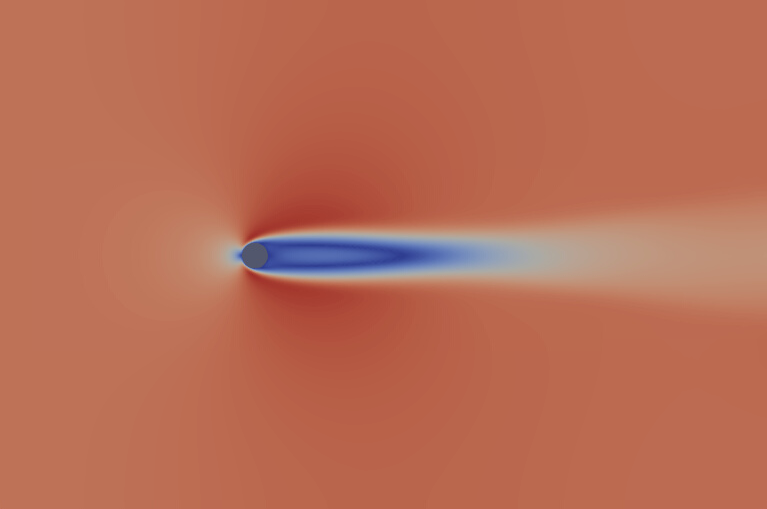} 
\end{subfigure}
\begin{subfigure}{.3\textwidth}
  \centering
  \includegraphics[width=\textwidth, trim={4cm 3cm 2cm 3cm}, clip]{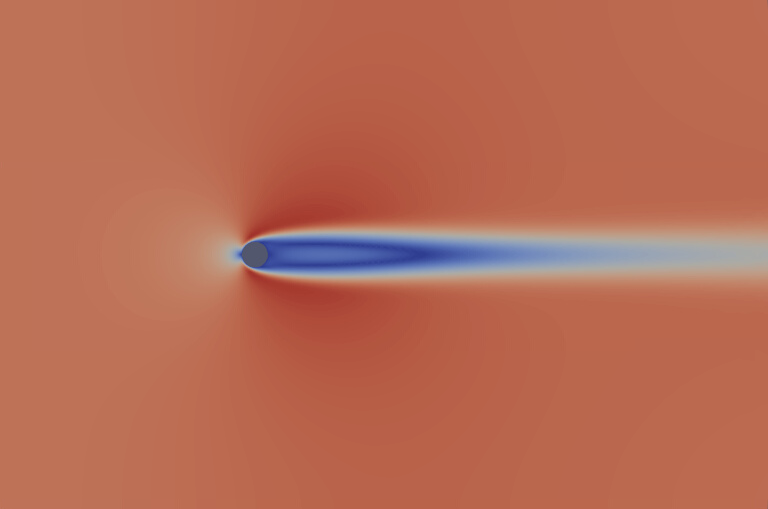} 
\end{subfigure}
\caption{Mean flow (velocity magnitude) $\bar \vecq = \langle \vecq(t) \rangle_t$ at the end of iterations 0 (uncontrolled, upper-left corner), 1, 2, 4, 6, 8 (last iteration, $\bar \vecq = \vecq_b$, lower-right corner). Color scaling is constant.}
\label{fig_meanflow}
\end{figure}

\paragraph{Recirculation bubble of the mean flow.}
The recirculation bubble of the mean flow is defined as the zone behind the cylinder, inside the contour delimited by $\vecv(x_1,x_2)=0$ on the mean flow $\bar\vecq= \langle \vecq(t) \rangle_t$ (figure \ref{fig_meanflow}). The recirculation bubble lengthens almost linearly throughout iterations, as the flow is being stabilized, until it finally reaches the base flow.

\begin{figure}
\centering
\begin{subfigure}{.3\textwidth}
  \centering
  \includegraphics[width=\textwidth, trim={5cm 3cm 2cm 3cm}, clip]{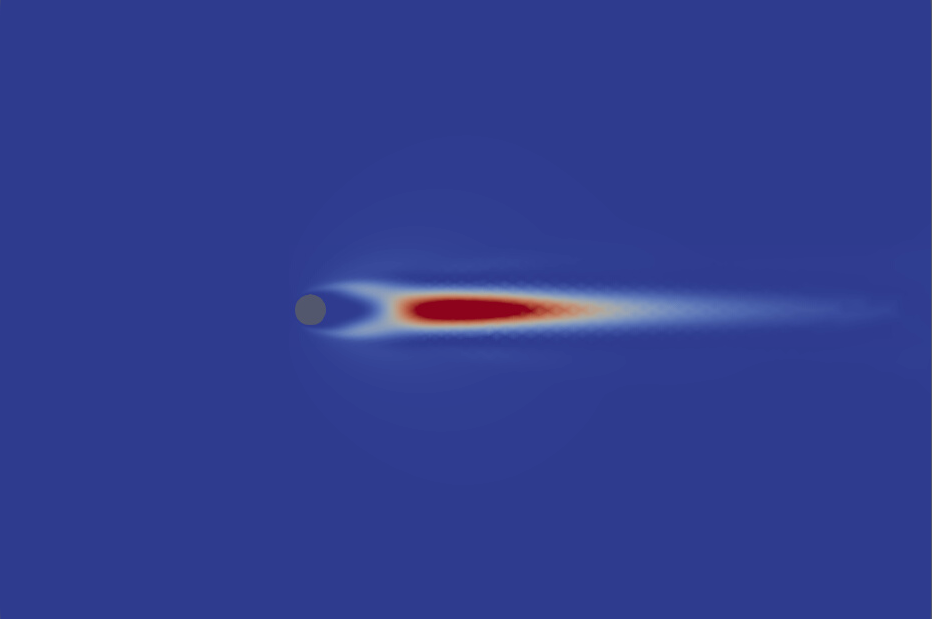} 
\end{subfigure}
\begin{subfigure}{.3\textwidth}
  \centering
  \includegraphics[width=\textwidth, trim={5cm 3cm 2cm 3cm}, clip]{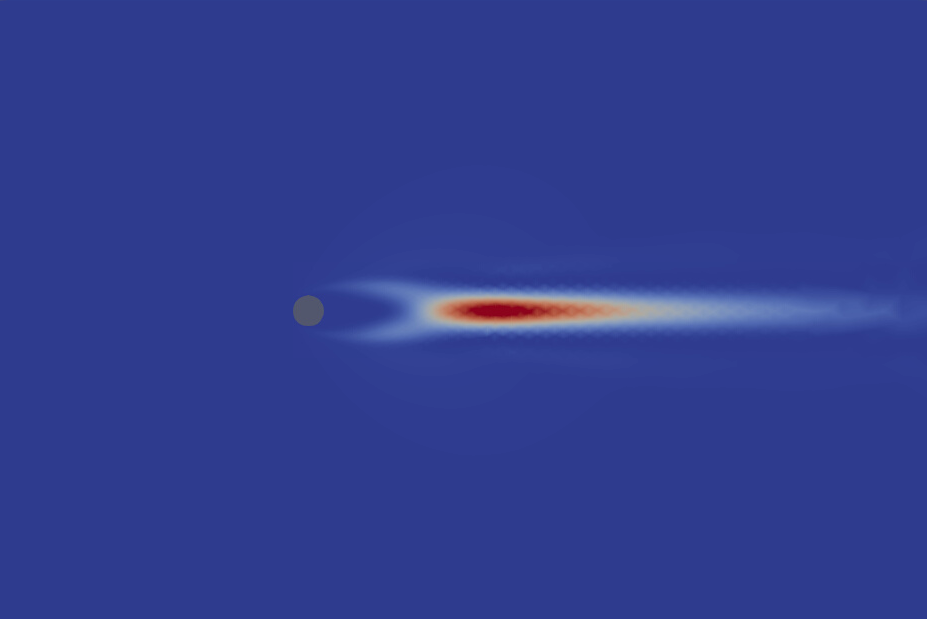} 
\end{subfigure}
\begin{subfigure}{.3\textwidth}
  \centering
  \includegraphics[width=\textwidth, trim={5cm 3cm 2cm 3cm}, clip]{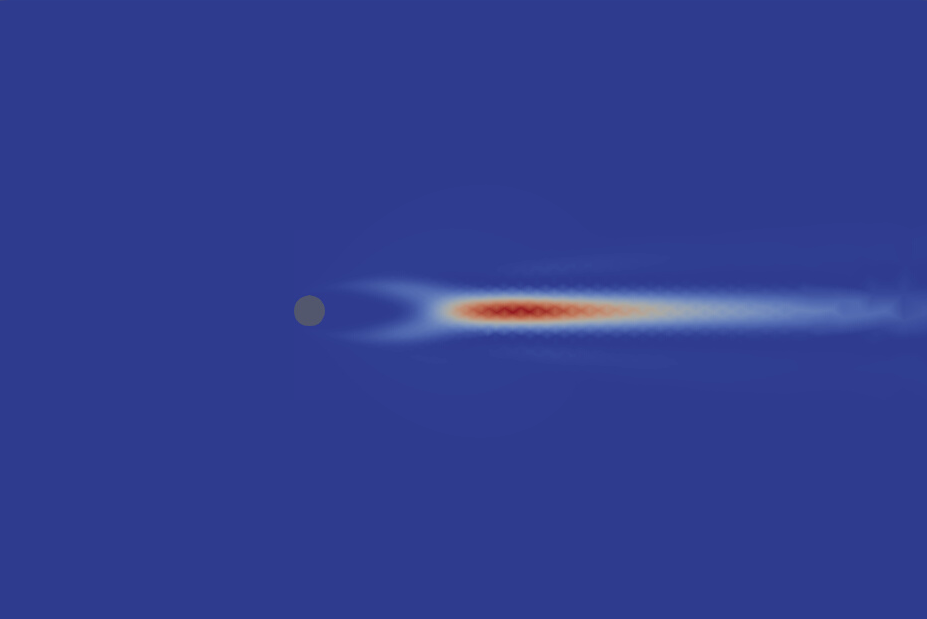} 
\end{subfigure}
\begin{subfigure}{.3\textwidth}
  \centering
  \includegraphics[width=\textwidth, trim={5cm 3cm 2cm 3cm}, clip]{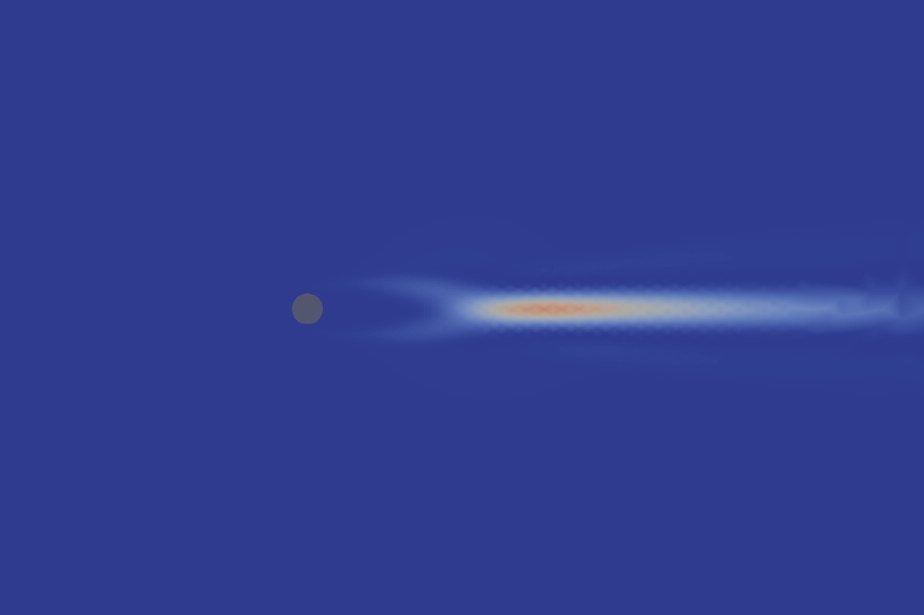} 
\end{subfigure}
\begin{subfigure}{.3\textwidth}
  \centering
  \includegraphics[width=\textwidth, trim={5cm 3cm 2cm 3cm}, clip]{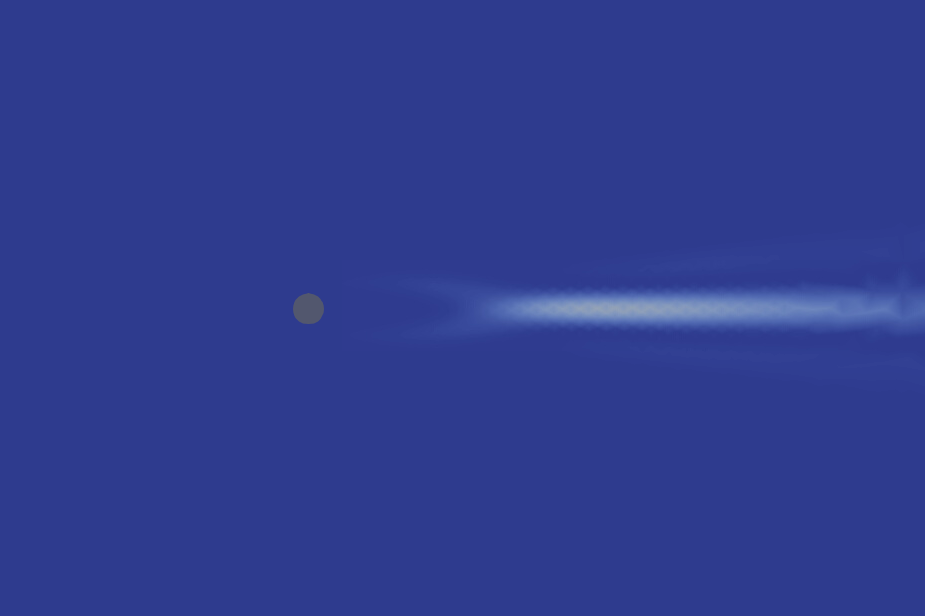} 
\end{subfigure}
\begin{subfigure}{.3\textwidth}
  \centering
  \includegraphics[width=\textwidth, trim={5cm 3cm 2cm 3cm}, clip]{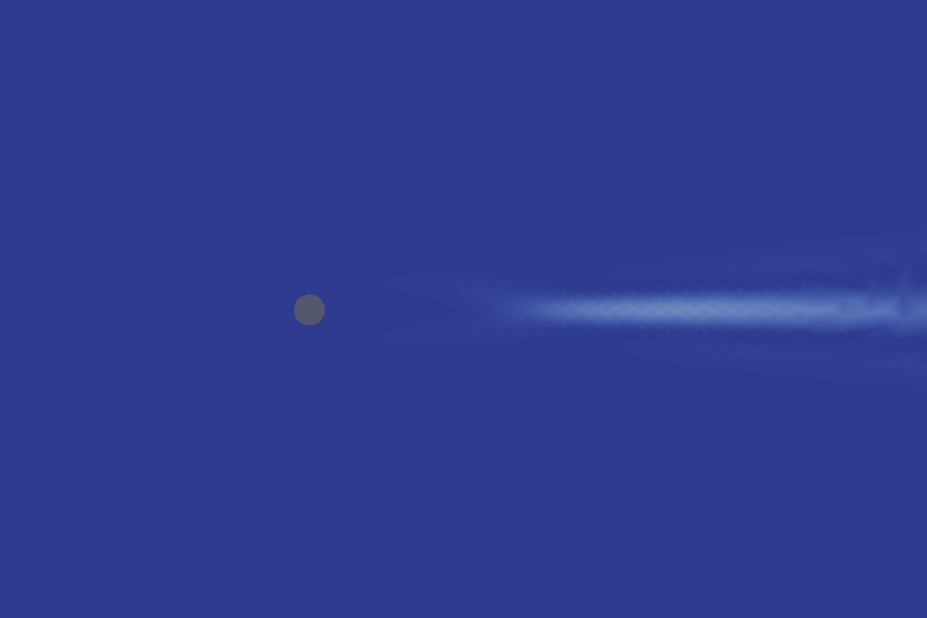} 
\end{subfigure}
\caption{Field $\epsilon(\bar \vecq)$ from iteration 0 (upper-left corner) to iteration 5 (lower-right corner). Color scaling is constant.}
\label{fig_epsilon}
\end{figure}

\paragraph{Spatial distribution of PKE.}
Throughout iterations, the spatial distribution of PKE evolves in a manner similar to the mean recirculation bubble. It is quantified with the mean flow deviation from the base flow, i.e. with the steady field $\vect{\epsilon}(\bar\vecq) = (\bar\vecq - \vecq_b)^T \cdot \mat{E} (\bar \vecq - \vecq_b)$ (figure \ref{fig_epsilon}).
It is notable from figure \ref{fig_epsilon} that the PKE peak is located far downstream in the wake, near the downstream limit of the mean recirculation bubble. The mean flow remains symmetric with respect to the $x_2$-axis, and the PKE is pushed downstream during iterations, with its peak value decreasing exponentially in amplitude.

%%%%%%%%%%%%%%%%%%%%%%%%%%%%%%%%%
\subsubsection{Frequency and nonlinearity analysis: spectrogram.} \label{sec_result_frequency}
\begin{figure}
\centering
\includegraphics[width=0.8\textwidth]{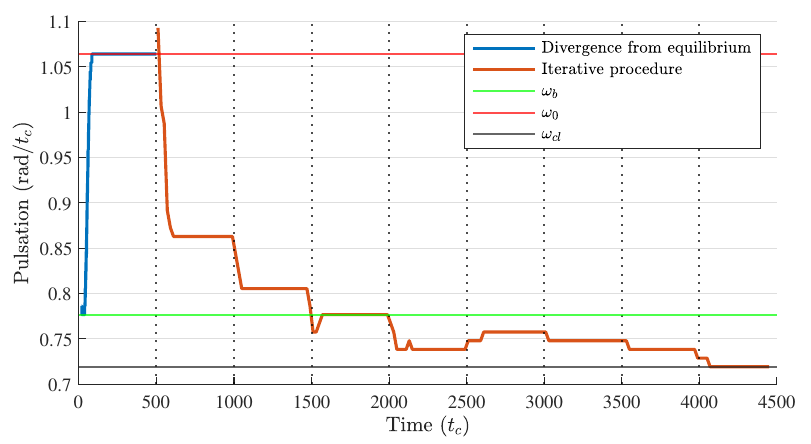}
\caption{Spectrogram of the feedback sensor signal (dominant frequency versus time) throughout the iterative procedure. The blue curve corresponds to the trajectory from the equilibrium to the natural attractor, while the red curve corresponds to the iterative procedure itself. Notable pulsations are marked by horizontal lines: $\omega_b$ in green, $\omega_0$ in red, and $\omega_{cl}$ in black.}
\label{fig_spectro_conv}
\end{figure}

\paragraph{Frequency of the flow.}
The attractor of the flow (unforced or controlled) stays periodic throughout iterations, which facilitates the analysis to be done with spectrograms: only one fundamental frequency and its harmonics are visible. 
The flow frequency throughout iterations can be extracted from the feedback probe. 
As mentioned in \S\ref{sec_unforced_flow}, the flow leaves the equilibrium with a pulsation $\omega_b = 0.779 \, \rad/t_c$ and settles on the natural attractor with a free pulsation $\omega_0 = 1.062 \, \rad/t_c$ after a continuous transition from one to the other. 
The evolution of the instantaneous fundamental frequency of the flow throughout the iterative process is displayed in figure \ref{fig_spectro_conv}: the blue curve is the divergence from the base flow (from $\omega_b$ (green line) to $\omega_0$ (red line)), and the red curve is the frequency throughout the iterative process (from $\omega_0$ to $\omega_{cl}$ (black line), which is defined in the paragraph below). 

In a very small amount of iterations ($\approx 3$), the frequency of the flow almost matches the frequency of the base flow, but the flow is not stabilized yet. Indeed, the frequency remains almost constant in subsequent oscillations, while the amplitude of all signals decreases. At the very last iterations, the frequency of the flow does not match $\omega_b$, which is expected. Indeed, the system at this point is equivalent to the closed-loop between the base flow $\Gb = \statespace{A_b}{B_b}{C_b}{0}$ and the last controller $K = \statespace{A_K}{B_K}{C_K}{0}$. The dynamic matrix of the closed-loop is: $\mat{\bar{A}} = \begin{bmatrix} \mat{A_b} & \mat{B_b} \mat{C_K} \\ \mat{B_K} \mat{C_b} & \mat{A_K} \end{bmatrix}$ and its singular mass matrix is: $\mat{\bar{E}}=\begin{bmatrix} \mat{E} & \mat{0}  \\ \mat{0}  & \mat{I} \end{bmatrix}$. The frequency $\omega_{cl}$ of the flow at the last iteration matches the least damped pole of this system, i.e. it is the imaginary part of the eigenvalue with largest real part, from the following generalized eigenvalue problem (singular, sparse, high-dimensional): $\lambda \in \mathbb{C}  \text{ s.t. }  \exists \vect{x} \neq \mat{0}: \mat{\bar{A}} \vect{x} = \lambda \mat{\bar{E}} \vect{x} $. 

\paragraph{Nonlinearity weakening throughout iterations.}
Another interesting observation is that nonlinearities are less and less active during iterations, symmetrically to the observations on the divergence from the base flow. Previously in \S\ref{sec_unforced_flow} (figure \ref{fig_spectro_div}), it was observed that nonlinearities appeared after a few time instants after perturbing the flow from its equilibrium, as higher-order harmonics in the frequency content of signals. When converging to the equilibrium with the iterative procedure, the observation is reversed (but not shown in figure \ref{fig_spectro_conv} for space reasons): the higher-order harmonics are very present in the first $3$ iterations, and they almost disappear in subsequent iterations. However, even with weaker nonlinearity, the stabilization of the flow still takes a moderate amount of additional iterations.

%%%%%%%%%%%%%%%%%%%%%%%%%%%%%%%%%
\subsubsection{Control law}\label{sec_results_control}
\paragraph{A piecewise-LTI control law.}
The control law produced by the iterative procedure is in essence piecewise-LTI: for a fixed iteration index, the total controller is a sum of LTI controllers, and the iteration index is a piecewise-constant function of time. The control law at any given time instant $t$ can be expressed as $K(t, s) = K_{i(t)}(s)$ where $i(t)$ is a piecewise-constant function of time. The control law can also be considered adaptive, in that the trajectory in phase space is not defined a priori; instead, at each iteration, controllers are synthesized to control a specific regime of the flow and reach a new, previously unknown regime.

It is notable that implementing the final controller directly from the limit cycle does not lead to stabilization of the flow, while the same controller implemented at the last iteration stabilizes the flow. First, it confirms that the control law produces a finite basin of attraction around the equilibrium, that does not encompass the natural limit cycle. Second, it indicates that the control law found here uses the variation in time of the flow to ultimately reach equilibrium, which is confirmed in the following. For the cylinder, several studies have reported full stabilization of the flow with single LTI controllers (finite basin of attraction encompassing the natural attractor, and no time variation of the control law) at $\Rey=100$, but they usually use strong model hypotheses such as ROMs from linearization or Galerkin projection (see e.g. \cite{illingworth2016,camarri2010feedback,jussiau2022learning}).

\begin{figure}
\centering
\includegraphics[width=0.7\textwidth]{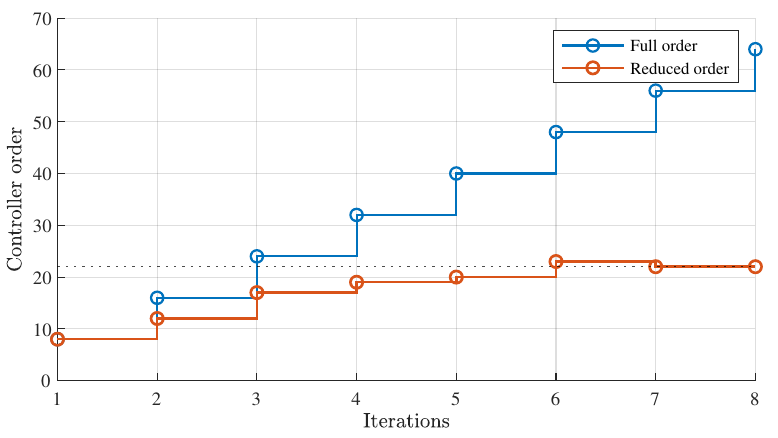}
\caption{Controller order throughout iterations, without reduction method (blue) and with balanced truncation (red).}
\label{fig_controller_order}
\end{figure}

\paragraph{Controller order reduction.}
Without controller order reduction, the order of the controller at iteration $i$ would be $i\times n_r$, while balanced truncation permits a controller order remaining almost constant throughout iterations, peaking at $n_K=23$, then settling down to $n_K=22$ (to be compared with a full-order of $64$ at the last iteration without balanced truncation in the process), as per figure \ref{fig_controller_order}. If needed, the reduction can also be performed with a maximum order constraint instead of a fixed HSV threshold $g_\sigma$, but there is a risk of neglecting critical dynamics for efficient control.

\begin{figure}
\centering
\includegraphics[width=0.7\textwidth]{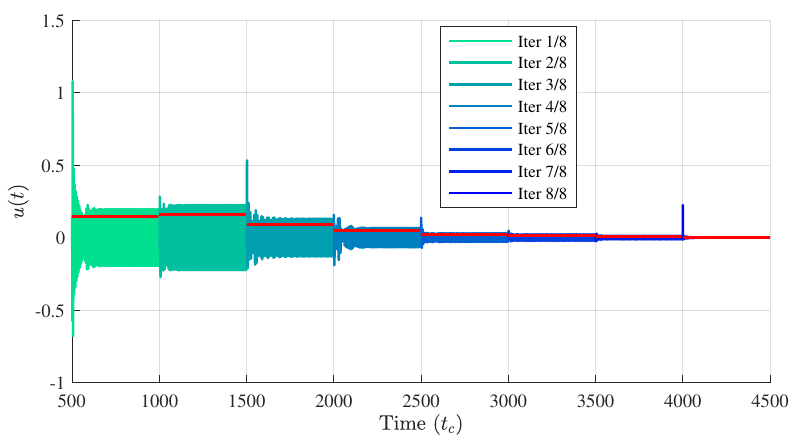}
\caption{Control input $u(t)$ throughout iterations. $\rms(u(t))$ is indicated by a red horizontal level at each iteration.}
\label{fig_u}
\end{figure}

\paragraph{Control input.}
One objective of the switching method from \S \ref{sec_controller_switching} is to induce small transients in the control input, and still manage to control the flow. The control input $u(t)$ throughout iterations is represented in figure \ref{fig_u}. As expected from the tuning of LQG control at the first iteration, there is large overshoot in the transient of the control input at time $t_1=500$.
In the following iterations, moderate transient in the control input is indeed observed, although the control input peaks more at some iterations (e.g. iterations $3, 8$).

A last observation is that stabilizing the flow does not require increasing control input power in general. Indeed, the control input is maximal during the first iterations and keeps reducing as iterations advance, to finally reach $u(t) \to 0$ on the equilibrium as $t \to \infty$. It underlines that the first iterations are the most critical from a control input point of view, and that constraints on the control input (i.e. on the controller gain) may be relaxed as the flow comes closer to equilibrium.

\paragraph{Independence of dynamical equilibrium from controller initial state.} \label{sec_independence}
On this configuration, the statistically-steady flow regime, reached after adding a controller in the loop, seems almost independent from the initial condition of the controller. Controller initialization only has an  impact on the transient before reaching said statistically-steady regime, which allows using any technique for switching, such as the ones evoked in \S\ref{sec_controller_switching} if the transient is not deemed to be an important factor. 
As the two-dimensional flow past a cylinder at $\Rey=100$ naturally exhibits only one attractor, this observation may be different on a more complex flow where several attractors exist simultaneously, e.g. the flow over an open cavity in \cite{bengana2019} for $4410 \leq \Rey \leq 4600$.

%%%%%%%%%%%%%%%%%%%%%%%%%%%%%%%%%%%%%%%%%%%
\section{Discussion}\label{sec_discussion}
\subsection{Modal analysis: mean flow and implicit models} \label{sec_implicit}
\subsubsection{Closed-loop identification and implicit flow model} 
\begin{figure}
\centering
\includegraphics[width=0.7\textwidth]{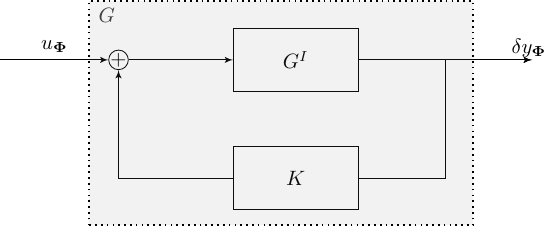}
\caption{Block diagram of the identified closed-loop system $G$ (mean transfer function) and the model of the flow alone $G^I$ (implicit model). The feedback of the system $G^I$ with the known controller $K$ produces the identified closed-loop $G$.}
\label{fig_implicit_model}
\end{figure}

At each iteration, whenever the flow is in feedback with the control law $K$, it lies on a dynamical equilibrium and we aim at identifying the mean transfer function $G$ for the following control step. The mean transfer function $G$ corresponds to a closed-loop system: it is the flow alone, in feedback with a known controller $K$. Therefore, one may remove the influence of the controller, to retrieve an LTI model of the flow itself, denoted $G^I$, as illustrated in figure \ref{fig_implicit_model}.
More specifically, we can write:
\begin{equation}
 G = \FB(G^I, K) =  \frac{G^I}{1 - G^I K} \, , 
\end{equation}
and one can then deduce the \emph{implicit} model of the flow alone $G^I$:
\begin{equation}
G^I =  \frac{G}{1 + GK} \, . 
\end{equation}
From the frequency response of $G$ and $K$, we compute the frequency response: ${G^I(\jw) = \frac{G(\jw)}{1 + G(\jw)K(\jw)}}$, and identify $G^I(s)$ as a low-order model. In practice, it proves satisfying to choose $G^I(s)$ with the same order as $G(s)$.

\subsubsection{Implicit flow model and mean flow model}
\begin{figure}
\centering
\includegraphics[width=0.7\textwidth]{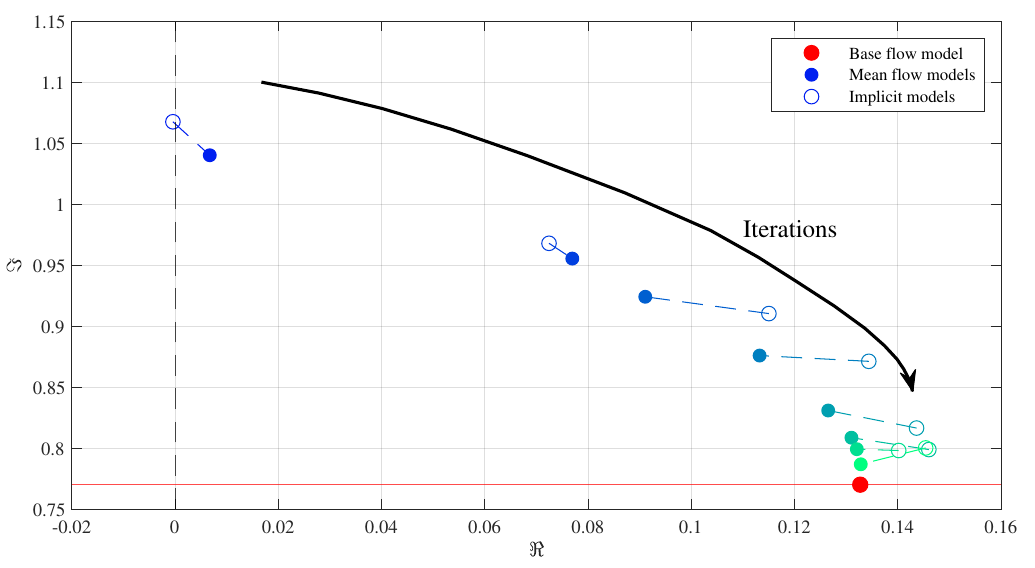}
\caption{Unstable pole of the implicit model $G^I_i$ (empty circles) and mean flow model $\bar{G}_i$ (filled circles) in complex plane, throughout iterations. The red dot is the unstable pole of the base flow model $\Gb$, with its frequency represented as the red line.}
\label{fig_pzmap_meanflow_vs_implicit}
\end{figure}

A question that might arise is the meaning of this implicit flow model. 
To that aim, we compare the implicit flow model to the mean flow model used in \cite{leclercq2019} for control purposes. The mean flow model corresponds to the linearization of the equations of the flow \eqref{eq_ns_q} around the mean flow $\bar{\vecq}$, and is denoted $\bar{G}$.
It is easily deduced that the two models are not strictly identical. At the first iteration, the implicit flow model is $G^I_0 = G_0$ (mean transfer function) because the controller in the loop is $K_0=0$, so its least stable poles are stable with very low damping (numerical artifact); and at the same time, the mean flow model $\bar{G}_0$ (figure \ref{fig_pzmap_meanflow_vs_implicit}) has an unstable pole. 

\paragraph{Displacement of the unstable pole in the complex plane.}
However, it can be checked that at each iteration, both models $G^I_i$ and $\bar{G}_i$ remain close together. The location of their respective unstable pole is tracked in figure \ref{fig_pzmap_meanflow_vs_implicit} (empty circles for $G^I_i$, filled circles for $\bar{G}_i$, linked by a dotted line for the same iteration). At the first iteration, the pole of $G^I_0$ has very low damping, while the pole of $\bar{G}_0$ is unequivocally unstable. As iterations progress, the poles drift together towards the unstable pole of the base flow (in red), while remaining close to each other at each iteration.

\begin{figure}
\centering
\includegraphics[width=0.7\textwidth]{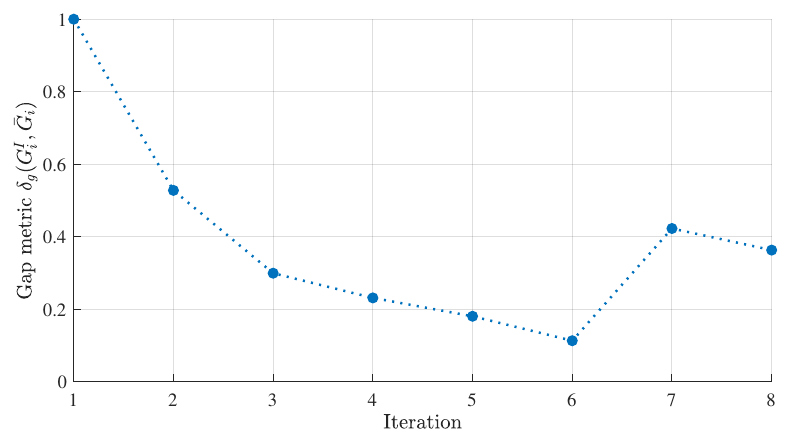}
\caption{Gap metric $\delta_g (G^I_i, \bar{G}_i )$ quantifying the resemblance from a closed-loop viewpoint of the mean flow model $\bar{G}_i$ and the implicit flow model $G^I_i$ throughout iterations.}
\label{fig_gapmetric_meanflow_vs_implicit}
\end{figure}

\paragraph{Closed-loop resemblance: the gap metric.}
Additionally, models are compared in figure \ref{fig_gapmetric_meanflow_vs_implicit} with the gap metric \citep[~Chap. 17]{zhoudoyle}, denoted $\delta_g(\cdot, \cdot) \in [0, 1]$. Note that a similar metric, known as the $\nu$-gap metric, was used in the past, in the context of robust control of flows in \cite{li2016feedback}.
For plants $G_1, G_2$ with normalized right coprime factorizations $G_1 = N_1 M_1^{-1}, G_2 = N_2 M_2^{-1}$, the gap metric is computed as:
\begin{equation}
\delta_g(G_1, G_2) = \max \left\{ \vec{\delta}(G_1, G_2), \vec{\delta}(G_2, G_1) \right\},
\end{equation}
where $\vec{\delta}(\cdot, \cdot)$ is the \emph{directed} gap metric that can be computed as follows:
\begin{equation}
\vec{\delta}(G_1, G_2) = 
\inf_{Q \in \RHinf} 
\norm{\begin{bmatrix}M_1 \\ N_1 \end{bmatrix} - 
\begin{bmatrix}M_2 \\ N_2 \end{bmatrix}Q}_\Hinf
,
\end{equation}
with $\RHinf$ the set of stable rational transfer functions.
The gap metric is a way to quantify the resemblance of models from a closed-loop perspective (instead of their open-loop behavior with the $\Hinf$ norm of the difference, which is infinite for unstable plants). Indeed, from \cite{zhoudoyle}, if the feedback between a plant $G$ and a controller $K$ is stable with given generalized stability margin:
\begin{equation} 
b_{G, K} = \norm{\begin{bmatrix} I \\ G \end{bmatrix} (I - K G)^{-1} \begin{bmatrix} K & I \end{bmatrix} }_\Hinf^{-1} = \norm{\begin{bmatrix} KS & S \\ T & GS \end{bmatrix}}_\Hinf^{-1}
,
\end{equation}
then any feedback between a plant $G'$ and the same controller controller $K$ is stable, if and only if:
\begin{equation}
\arcsin b_{G, K}  \geq \arcsin \delta_g(G', G),
\label{eq_gapmetric_arcsin}
\end{equation}
which provides a similarity metric between plants in closed-loop with an identical controller. More specifically, from \eqref{eq_gapmetric_arcsin}, if $\delta_g \sim 0$ then two plants are likely to be stabilized by the same set of controllers, and conversely for $\delta_g \sim 1$.

Now, regarding the implicit model $G^I$ and the mean flow model $\bar{G}$, what appears in figure \ref{fig_gapmetric_meanflow_vs_implicit} strengthens the observation from the previous paragraph: it not only confirms that models remain close, but also that they become closer together as the flow approaches the equilibrium. At the last two iterations, the jump in gap metric is unexplained, but the level of gap metric stays low nonetheless.

\begin{figure}
\centering
\includegraphics[height=6.2cm]{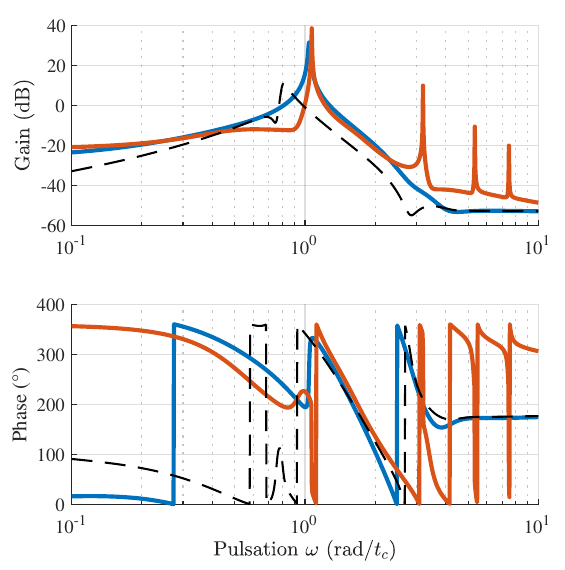}
\includegraphics[height=6.2cm]{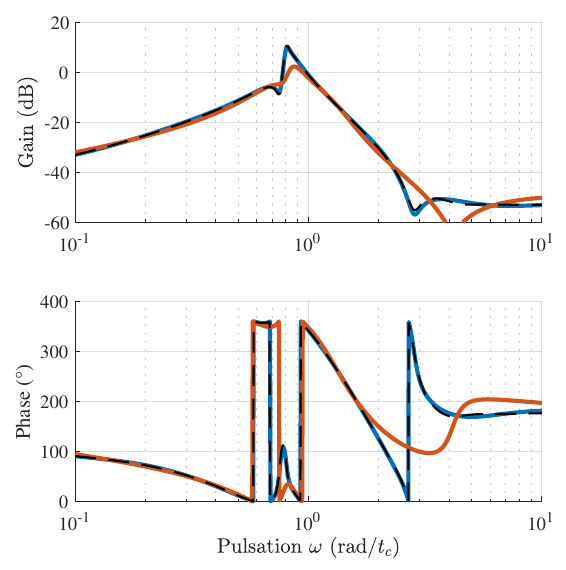}
\caption{Bode diagrams of mean flow $\bar{G}$ (blue) and implicit $G^I$ (red) models. The left panel is at iteration $1$ (unactuated flow), while the right panel is at the iteration $8$ (last). The base flow model $\Gb$ is in dotted black.}
\label{fig_bode_meanflow_vs_implicit}
\end{figure}

\paragraph{Bode diagrams.}
Finally, we also depict the Bode diagram of the implicit flow model and the mean flow model in figure \ref{fig_bode_meanflow_vs_implicit}. While the resemblance of $\bar{G}_0, G^I_0$ is not obvious (slightly different frequency of the resonant/unstable pole, additional harmonics in the implicit mode explained in \cite{leclercq2023}), both $\bar{G}_7$ and $G^I_7$ are very close together, and close to the base flow model $\Gb$ (although the implicit model pole has larger damping).

It would be interesting to address more intricate flow regimes, such as the quasiperiodic flow over an open cavity \citep{leclercq2019}, that displays several unstable poles in its mean flow model, and investigate whether the mean flow model and the implicit flow model have common features in general.

\subsubsection{Nonlinear relaxation of poles}
\begin{figure}
\centering
\includegraphics[width=0.9\textwidth]{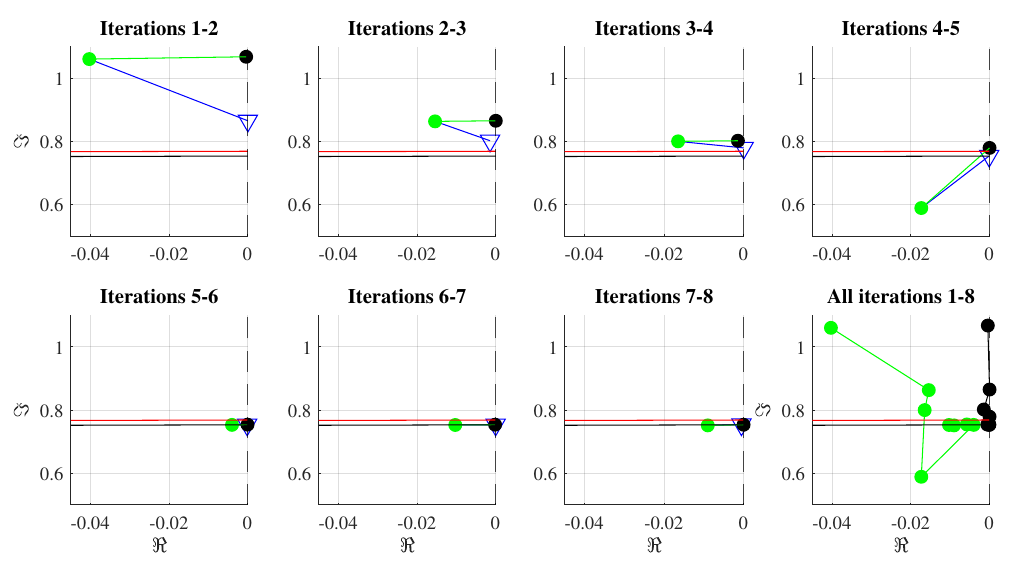}
\caption{Nonlinear relaxation of resonant pole associated to the unique fundamental frequency, throughout iterations. The resonant pole identified from the mean transfer function $G_i$ (black dot) is displaced into the stable plane (in green) by the controller $K^+_i$. A nonlinear transient regime leading to a new dynamical equilibrium shifts the stabilized pole to the imaginary axis again (in blue). The red line is the pulsation of the base flow $\omega_b$, the black line is the pulsation of the final closed-loop $\omega_{cl}$. In the last panel, the route of the stabilized poles is indicated in green.}
\label{fig_relaxation}
\end{figure}

At each iteration, the mean transfer function is supposed to exhibit poles on the imaginary axis $i\mathbb{R}$, although the identification of $G_i$ places the poles in the stable half-plane. The aim of the controllers $K_i^+$ is to damp the poles of $G_i$, farther away from the imaginary axis. 
After a transient regime where the flow shifts in phase space and reaches a new dynamical equilibrium, a new mean transfer function $G_{i+1}$ can be identified as well, and also has poles near the imaginary axis. 
Therefore, poles of $G_i$ are first damped by an LTI controller $K_i^+$, then after a transient regime, are relaxed and rejoin the vicinity of the imaginary axis again, which is illustrated in figure \ref{fig_relaxation}. 
This relaxation resembles a similar phenomenon, referred to as \emph{nonlinear relaxation}, in \cite{leclercq2019}. 
Below, we display the movement of poles, from $i\mathbb{R}$ to the stable plane as they are damped by the controller, then back to $i\mathbb{R}$ due to the nonlinear transient. 
The same observation as in figure \ref{fig_spectro_conv} can be made, in that the frequency of the oscillation quickly becomes constant, as the imaginary part of the pole approaches the black line at $\omega_{cl}$.

\subsection{Unmodeled phenomena and gaps to cover for experiments}
Oscillator flows are typically considered as being dominated by self-sustained oscillations and mostly insensitive to perturbations (in a broad sense: numerical or experimental noise, sometimes referred to as {state noise}), in contrast to noise amplifiers. 
However, in the context of periodic oscillator flows, it was shown in \cite{bagheri2014}, that a noisy limit cycle is defined by two time scales: its natural time scale $T$, and a correlation time $t_d$ characterizing \emph{phase diffusion}, which may cause oscillations to become uncorrelated after a certain duration. Depending on the \emph{quality factor} $\mathcal{Q}=2\pi \frac{t_d}{T}$, the impact of noise may be more or less important. 
Additionally, it is shown that eigenvalues of the Koopman operator (which are the poles of the mean transfer function \citep{leclercq2023}) are expected to be damped instead of purely imaginary. 
For the transfer function estimation in the presence of a noisy limit cycle, more sophisticated estimation techniques should be used, such as the ones described in \cite{van2012subspace} for identification, or \cite{bendat2011random} for random noise processing. 

Regarding actuation and sensing, we have assumed that both the actuator and sensor could be modeled in a linear fashion, although this is transparent from the data viewpoint. In the future, attention should be paid to the impact of actuator saturations or rate saturations. Also, both the actuator or sensor may be corrupted by noise, to which system identification is naturally robust, but the impact should be assessed nonetheless. For the sensor specifically, using a sensor in the wake in this study may not be limiting, as the procedure is likely to be qualitatively comparable with a wall-mounted sensor.

%%%%%%%%%%%%%%%%%%%%%%%%%%%%%%%%%%%%%%%%%%%%%%%%%%%%%
\section{Conclusion and perspectives}\label{sec_conclusion}
 In this article, we have proposed an automated iterative methodology for the complete stabilization of the flow past a cylinder at $\Rey=100$ on its unknown natural equilibrium, using solely input-output data and linear time-invariant (LTI) controllers, to produce a piecewise-LTI adaptive control law of low complexity.
The objective of the approach is to drive the flow from its fully-developed regime of vortex-shedding, to its natural unstable equilibrium stabilized in closed-loop in a linear sense. This is achieved in simulation with a sequence of $8$ LTI controllers of moderate order, synthesized using models inferred from input-output data.

  The methodology builds upon a previous linear iterative method for the closed-loop control of oscillator flows, proposed in \cite{leclercq2019}. The principle is to solve this complex nonlinear control problem iteratively: instead of addressing stabilization of a high-dimensional nonlinear system, the problem is solved as a sequence of low-order linear approximations of the problem, whose resolution iteratively drive the system closer and closer to equilibrium.
Although this study demonstrated its capability to fully stabilize the quasiperiodic flow over an open cavity, it is not entirely applicable to an automated experimental setup: it relies on the linearization of the equations about the mean state; it uses structured $\Hinf$ synthesis that is arguably hard to automate reliably; and the order of the controller is increasing linearly with the iterations. 
In the present study, using the same principle, we target full automation of the procedure and rely only on input-output data for experimental compatibility. The modeling of the flow with an LTI plant is justified by the \emph{mean resolvent} framework \citep{leclercq2023}, providing keys for estimating the mean transfer with measurable input-output data only; controller synthesis is addressed with the LQG framework; and the order of the controllers is managed with online balanced truncation.

The main result is that each new controller disrupts the current attractor, leading to a new dynamical equilibrium with lower perturbation kinetic energy, constituting an adaptive piecewise-LTI control law. This iterative process ultimately drives the flow to the natural equilibrium, stabilized in closed-loop. The key difference from the previous study \cite{leclercq2019} is that our approach does not assume prior knowledge of the equations nor the mean flow, and may operate with minimal human supervision. In simulation, the whole method is also computationally inexpensive: it only requires forward-time simulations of the actuated flow, and low-dimensional numerical algebra.

Several investigations should be conducted in the simulation in order to study the following points.
Firstly, for model-based LTI control strategies, the mean transfer function is, as an \emph{average} model, the most appropriate model choice for synthesis. 
The optimal class of input signals to obtain the frequency response of the mean transfer function efficiently \emph{in practice} (i.e. requiring the least amount of simulation or experimental time) is still an open question.
Also, to enhance flow control efficiency, the focus should not be directed towards extensively polishing the identification process, but rather towards refining the synthesis method. Although the LQG framework has demonstrated effectiveness in the low-gain limit and is easily automated, alternative model-based control strategies could be explored, possibly incorporating constraints : typically, MPC (as in \cite{arbabi2018}) with iterative model adaptation would naturally handle actuator saturations.

However, there is little hope that energy decrease can be predicted and optimized using solely linear criteria, as non-linear effects are not incorporated in the mean resolvent framework. Nevertheless, model-based LTI methods provide a simple yet efficient framework for building control strategies, even based on data, and they may be more phyiscally-grounded than the model-free control methods proposed by Reinforcement Learning (RL).
In this sense, we would argue that model-based LTI strategies could be fine-tuned with input-output data obtained directly from the nonlinear system, aligning with the approach in \cite{jussiau2022learning}.
As a final objective, the scope of this work suggests trying to implement the method in an experiment. While the approach is theoretically fully data-based, some aspects of the method will still require being addressed in a real-life setup, such as sensor and actuator noise, saturations and rate saturations, three-dimensionality or turbulence.

\backsection[Funding]{
This research was funded by ONERA - The French Aerospace Lab. For the purpose of Open Access, a CC-BY 4.0 public copyright licence (https://creativecommons.org/licenses/by/4.0/) has been applied by the authors to the present document and will be applied to all subsequent versions up to the Author Accepted Manuscript arising from this submission. 
}

\backsection[Declaration of interests]{The authors report no conflict of interest.}

%%%%%%%%%%%%%%%%%%%%%%%%%%%%%%%%%%%%%%
\appendix

\section{Mean transfer function estimation in practice}\label{appendix_identification}
We recall that the objective from \S \ref{sec_mean_freqresp} is to estimate the mean frequency response $H_0(\jw)$ from input-output data, for which some equations are reproduced below. For a multisine input $u_{\mathbf{\vect{\Phi}}}(t)$ at frequencies $k\omega_u$ and a measured output $y_{\mathbf{\vect{\Phi}}}(t)$, the Fourier coefficients can be found as harmonic averages:
\begin{equation}
\left\{
\begin{aligned} 
& \widehat{u}_{\mathbf{\vect{\Phi}}}(k\omega_u) = \lim_{T'\to \infty} \frac{1}{T'} \int_{0}^{T'} u_{\mathbf{\vect{\Phi}}}(t) e^{-\jwu t} \, \d t, \\
& \widehat{y}_{\mathbf{\vect{\Phi}}}(k\omega_u) = \lim_{T'\to \infty} \frac{1}{T'} \int_{0}^{T'} y_{\mathbf{\vect{\Phi}}}(t) e^{-\jwu t} \, \d t = \widehat{\delta y}_{\mathbf{\vect{\Phi}}}(k\omega_u).
\end{aligned}
\right.
\label{eq_harmonic_average_appendix}
\end{equation}
Then, the frequency response depending on the phase $\mathbf{\vect{\Phi}}$ of the input, may be computed as a ratio of Fourier coefficients:
\begin{equation}\label{eq_fourier_ratio_appendix}
H_{\mathbf{\vect{\Phi}}}(\jwu)  = \frac{\widehat{\delta y}_{\mathbf{\vect{\Phi}}}(k\omega_u)}{\widehat{u}_{\mathbf{\vect{\Phi}}}(k\omega_u) },
\end{equation}
which is such that $\Ep_{\mathbf{\vect{\Phi}}}(H_{\mathbf{\vect{\Phi}}}(\jwu))  = H_0(\jwu)$.
Below, we describe how the input signals $u_{\mathbf{\vect{\Phi}}}(t)$ are designed, how harmonic averages from \eqref{eq_harmonic_average_appendix} (initially in \eqref{eq_harmonic_average}) are approximated with Discrete Fourier Transforms (DFTs) in practice, and how the sample mean of $H_{\mathbf{\vect{\Phi}}}(\jwu)$ is computed.

\subsection{Design of numerical experiments and Discrete Fourier Transforms}
\begin{figure}
\centering
\includegraphics[width=0.8\textwidth]{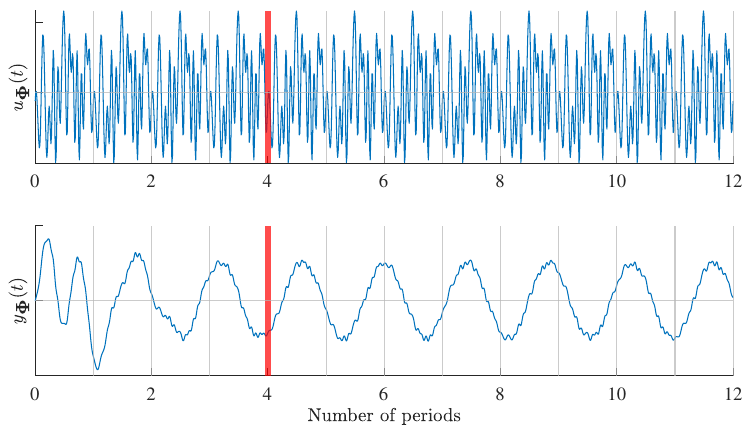}
\caption{Design of multisines -- Synthetic data. In this illustration, 12 periods of a periodic input $u_{\mathbf{\vect{\Phi}}}(t)$ are represented, along with the corresponding output $y_{\mathbf{\vect{\Phi}}}(t)$. The first $P_{tr}=4$ periods of both the input and the output are discarded, for containing a transient regime due to damped Floquet modes \citep{leclercq2023}. To mitigate the quasiperiodicity of the remaining portion of $y_{\mathbf{\vect{\Phi}}}(t)$, a Hann window is used when computing both DFTs.}
\label{fig_multisine_experiment}
\end{figure}

In practice, timeseries are extracted with time step $\Delta t$ and finite duration $T$ to be defined, and the computations are done in sampled time.
For the harmonic averages in \eqref{eq_harmonic_average} to be approximated with finite horizon $T$, the transient is discarded and a Discrete Fourier Transform (DFT) is computed with several adjustments. 
Indeed, when injecting the input $u_{\mathbf{\vect{\Phi}}}$, the measurement $y_{\mathbf{\vect{\Phi}}}$ undergoes a transient regime as per \cite{leclercq2023}, containing the contribution of damped Floquet modes, before settling in a statistically-steady regime, represented in figure \ref{fig_multisine_experiment}.
The contribution of the transient to the harmonic average \eqref{eq_harmonic_average} vanishes for $T\to\infty$ ;
in practice, it is better to suppress it from the dataset for an estimation based on a finite time window, using the DFT.  Therefore, the beginning of the experiment, corresponding to the first $P_{tr}$ periods of the signal $u_{\mathbf{\vect{\Phi}}}(t)$, is discarded.  
Also, as the DFT assumes periodicity of the signals, which is not the case for $y_{\mathbf{\vect{\Phi}}}(t)$, a Hann window is used. 
Accounting for the $P_{tr}$ discarded periods, the total duration of the experiment is $(P_{tr} + P) \frac{2\pi}{\omega_u}$ with $P_{tr}, P \in \mathbb{N}$, but a subset of length $T=P\frac{2\pi}{\omega_u}$ is used for the estimation.
The subsequent DFT resolution is by definition $\Delta \omega = \frac{2\pi}{T}$, such that $\omega_u = P \Delta \omega$, i.e. nonzero contributions in the input are retrieved every $P$ point in the DFT.
In practice, $P_{tr}=P=4$ are fixed, and the number of realizations $M$ for the ensemble average is chosen below.

\subsection{Convergence of the sample mean and the time-invariance hypothesis}\label{appendix_convergence}
The sample mean $\bar H(\jw)$ of $H_{\mathbf{\vect{\Phi}}}(\jwu)$ is computed by performing $M$ independent simulations of the nonlinear system with different realizations of the input $u_{\mathbf{\vect{\Phi}}}(t)$, from the same limit cycle phase $\phi$. The output $y_{\mathbf{\vect{\Phi}}}(t)$ of each simulation is gathered for the frequency response computation.
We perform a total of $\widehat{M}=16$ experiments. 
For each experiment $m\in [1, \widehat{M}]$, we can compute the frequency response $H_{\mathbf{\vect{\Phi}}}^m(\jw)$. 
Now, for efficient estimation in practice, we wish to be able to reduce $\widehat{M}$: we would like to find a lower value $M \leq \widehat{M}$ such that the sample mean depending on $M$, defined below, is \emph{converged}: 
\begin{equation}
\bar H_{\mathbf{\vect{\Phi}}}(\jw; M) = \frac{1}{M} \sum_{m=1}^{M} H_{\mathbf{\vect{\Phi}}}^m(\jw).
\end{equation}  
For that purpose, the reference is taken as $\bar H(\jw) = \bar H_{\mathbf{\vect{\Phi}}}(\jw; \widehat{M})$, which shows reasonable in practice since the value of $\bar H_{\mathbf{\vect{\Phi}}}(\jw; \widehat{M})$ is almost constant for $M \in [12, 16]$.

First, we consider that the sample mean is converged with respect to $M$ at a given frequency when $\sqrt{\Var \bar H_{\mathbf{\vect{\Phi}}}(\jw; M)} \ll | \bar H(\jw) |$. Since $\Var \bar H_{\mathbf{\vect{\Phi}}}(\jw; M) = \frac{1}{M} \Var H_{\mathbf{\vect{\Phi}}}(\jw)$, the previous condition translates into:
\begin{equation}
\zeta(\jw) = \frac{\sqrt{\Var H_{\mathbf{\vect{\Phi}}}(\jw)}}{| \bar H(\jw) |} \ll \sqrt{M},
\label{eq_zeta}
\end{equation}
In the above expression, the variance of the frequency response itself $\Var H_{\mathbf{\vect{\Phi}}}(\jw)$ is estimated once with the full data $M=\widehat{M}$. 
Note that the formulation \eqref{eq_zeta} is similar to that of \cite{leclercq2023} with the quantity $\eta(\jw; u)$ that quantifies the variation of the transfer with respect to the limit cycle phase $\phi$ (for a given input signal $u$). Here, $\phi$ is kept constant (signals are injected at the same phase of the limit cycle for every realization), but the phase $\mathbf{\vect{\Phi}}$ of the input signal itself is varied. %, which is quantified through $\zeta(\jw)$.

\begin{figure}
\centering
\includegraphics[width=0.8\textwidth]{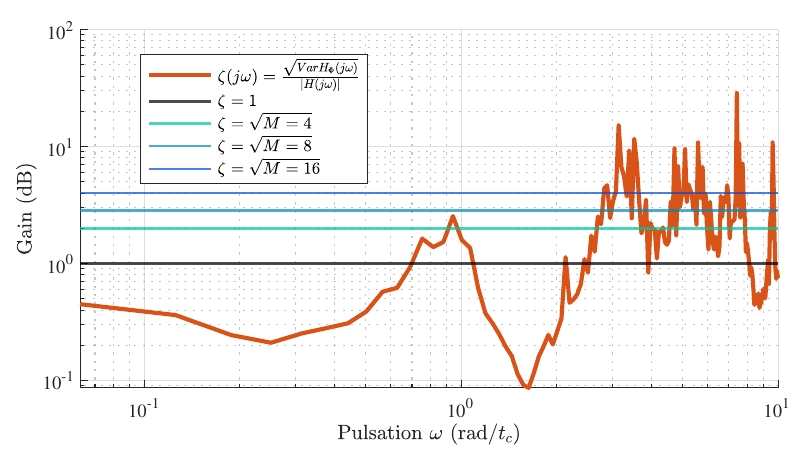}
\caption{Comparison of $\zeta(\jw)$ with $\zeta=1$ and $\sqrt M$ for $M=4, 8, 16$. It is notable that the mean transfer at $M=4$ cannot be considered entirely converged for $\omega\approx 0.9$ or $\omega \geq 3$. In turn, models presented in the paper have low reliability for $\omega \geq 3$.} 
\label{fig_variance_freqresp_M}
\end{figure}

% zeta=sqrtM : indicates whether the mean is converged
In figure \ref{fig_variance_freqresp_M}, we show the estimation of $\Var H_{\mathbf{\vect{\Phi}}} (\jw)$ on the whole pulsation range, and horizontal lines as $\sqrt{M}$ for $M=4, 8, 16$. In this situation, it appears that $M=4$ offers a fair estimation of the mean with low computational complexity. Although there seem to miss samples for the mean to be converged at high frequency ($\omega \geq 3$), a significant part of the low-frequency content indeed lies beyond the horizontal line $\zeta=\sqrt{4}=2$. 

% zeta=1 : indicates whether the TI hypothesis is valid
Besides, the horizontal line $\zeta=1$ in figure \ref{fig_variance_freqresp_M} conveys important information about the relevance of the time-invariant approximation of the flow.
Indeed, the samples of $H_{\mathbf{\vect{\Phi}}}(\jw)$ are distributed around their mean $\bar H(\jw)$ due to the time-dependence of the unforced flow \citep{leclercq2023}. When approximating the frequency response with its mean, representing a time-invariant model, it is important that $\bar H(\jw)$ conveys sufficient information. Here, $\zeta(\jw) \ll 1$ indicates that the mean transfer function is a satisfactory time-invariant approximation of the model at said frequency and for a given input signal (i.e., the time-varying effects are low), and conversely for $\zeta(\jw) \gg 1$. 
The limit $\zeta=1$ is indicated as a black horizontal line in figure \ref{fig_variance_freqresp_M} and it may be observed that the time-invariance hypothesis holds well in the range $\omega \in [0,2.5] \, \rad/t_c$, except in the vicinity of the resonance at $\omega \approx 1 \, \rad/t_c$. Similarly to \cite{leclercq2023}, the time-invariance hypothesis is poorer at higher frequency.

\section{Subspace identification method: FORSE} \label{appendix_forse}
The Frequency Observability Range Space Extraction (FORSE, \cite{forse}) estimates a discrete-time state-space representation in two distinct steps, from the frequency response $\{H(\jw_i)\}_{i=1}^M$. 
We denote $\Delta t$ the sampling time of the discrete-time SISO system $(\mat{\hat{A}}, \mat{\hat{B}}, \mat{\hat{C}}, \mat{\hat{D}})$ estimated by the algorithm.
We start by constructing a matrix containing the shifted frequency response: 
\begin{equation}
\Y = \begin{bmatrix}  H(\jw_1) & \hdots & H(\jw_M) \\ 
H(\jw_1) e^{\jw_1 \Delta t} & \hdots & H(\jw_M) e^{\jw_M \Delta t} \\ 
\vdots & \vdots & \vdots \\ 
 H(\jw_1)e^{\jw_1 (q-1) \Delta t}  & \hdots & H(\jw_M)e^{\jw_M (q-1) \Delta t} \end{bmatrix}
 ,
\end{equation}
where $q \in \mathbb{N}$ is a parameter of the identification algorithm.
When working with a frequency response directly, the influence of input and output is included in the matrix $\Y \in \mathbb{C}^{q\times M}$, and an impulse input matrix $\U \in \mathbb{C}^{q\times M}$ needs to be constructed, as follows:
\begin{equation}
\U = \begin{bmatrix}  1 & \hdots & 1 \\ 
\vdots & \vdots & \vdots \\ 
 e^{\jw_1 (q-1) \Delta t}  & \hdots & e^{\jw_1 (q-1) \Delta t} \end{bmatrix}
 .
\end{equation}
At this point, it is possible to weight the frequency response with a positive-definite diagonal matrix $\mat{R}\in\mathbb{R}^{M\times M}$ in order to focus on particular frequency domains (e.g. resonant modes). We denote the real part with $\Re$, the Moore-Penrose pseudo-inverse with the superscript $\dag$, and construct the following real matrix $\vect{\mathcal{H}}$, on which we subsequently perform a Singular Value Decomposition (SVD):
\begin{align}
\vect{\mathcal{H}} &= \Re(\Y \vect{R} \Y^*) - \Re(\Y \vect{R} \U^*)\Re(\U \vect{R} \U^*)^\dag \Re(\Y \vect{R} \U^*)^T \nonumber \\
 &= \begin{bmatrix} \vect{\Psi}_q & \vect{\tilde{\Psi}}_q \end{bmatrix}   \begin{bmatrix} \vect{\Sigma} & 0 \\ 0 & \vect{\tilde{\Sigma}} \end{bmatrix}  \begin{bmatrix} \vect{\Psi}_q \\ \vect{\tilde{\Psi}}_q \end{bmatrix}.
\end{align}
The $n_r$ largest singular values are stored in the diagonal matrix $\vect{\Sigma}$ associated to the singular vectors $\vect{\Psi}_q$, and the rest is discarded, dictating the precision of the ROM. Then, we partition $\vect{\Psi}_q$ as rows $\vect{\psi}_i \in \mathbb{C}^{1 \times n_r}$, and construct submatrices $\vect{\Psi}_{q-1},  \vect{\hat{\Psi}}_{q-1}$ such that:
\begin{equation}
\vect{\Psi}_q = \begin{bmatrix} \vect{\psi}_0 \\ \vect{\psi}_1 \\ \vdots \\ \vect{\psi}_{q-2} \\  \vect{\psi}_{q-1} \end{bmatrix}
= \begin{bmatrix} \vect{\Psi}_{q-1} \\  \vect{\psi}_{q-1}  \end{bmatrix}
= \begin{bmatrix} \vect{\psi}_0 \\  \vect{\hat{\Psi}}_{q-1} \end{bmatrix}
.
\end{equation}
Finally, we estimate the discrete-time dynamics $\mat{\hat{A}}$ and the measurement matrix $\mat{\hat{C}}$ as follows:
\begin{equation}
\mat{\hat{A}} = \vect{\Psi}^\dag_{q-1}  \vect{\hat{\Psi}}_{q-1},   \qquad   \mat{\hat{C}} = \vect{\psi}_0.
\label{eq_forse_pseudoinverse}
\end{equation}
If we wish to introduce stability constraints on the system to be identified, the aforementioned pseudo-inverse operation \eqref{eq_forse_pseudoinverse} is augmented with Linear Matrix Inequalities (LMIs) on $\mat{\hat{A}}$ \citep{fabricelmi}.
After $\mat{\hat{A}}, \mat{\hat{C}}$ have been estimated, the computation of the frequency response from the state-space realization is linear in $\mat{\hat{B}}$ and $\mat{\hat{D}}$ at each frequency:
\begin{equation}
H(\jw_k) = \mat{\hat{C}}(e^{-\jw_k \Delta t}\mat{I} - \mat{\hat{A}})^{-1}\mat{\hat{B}} + \mat{\hat{D}} = \vect{\mathcal{P}}(\jw_k) \begin{bmatrix} \mat{B} \\ \mat{D} \end{bmatrix}.
\end{equation}
Finding estimates of $\mat{\hat{B}}, \mat{\hat{D}}$ is then reformulated as a linear least-squares problem:
\begin{equation}
\mat{\hat{B}}, \mat{\hat{D}}=\arg\min_{\mat{B}, \mat{D}}   \norm{ \begin{bmatrix} H(\jw_1) \\ \vdots \\ H(\jw_M) \end{bmatrix}    -   
  \begin{bmatrix} \vect{\mathcal{P}}(\jw_1) \\ \vdots \\ \vect{\mathcal{P}}(\jw_M) \end{bmatrix}     \begin{bmatrix} \mat{B} \\ \mat{D} \end{bmatrix}}_{F}^2.
\end{equation}

%%%%%%%%%%%%%%%%%%%%%%%%%%%%%%%%%%%%%%
\bibliographystyle{jfm} 
\bibliography{mybibfile}

\begin{thebibliography}{87}
\expandafter\ifx\csname natexlab\endcsname\relax\def\natexlab#1{#1}\fi
\def\au#1{#1} \def\ed#1{#1} \def\yr#1{#1}\def\at#1{#1}\def\jt#1{\textit{#1}}
  \def\bt#1{#1}\def\bvol#1{\textbf{#1}} \def\vol#1{#1} \def\pg#1{#1}
  \def\publ#1{#1}\def\arxiv#1{#1}\def\org#1{#1}\def\st#1{\textit{#1}}

\bibitem[Aleksi{\'c} {\em et~al.\/}(2010)Aleksi{\'c}, Luchtenburg, King, Noack
  \& Pfeifer]{king2010}
{\sc \au{Aleksi{\'c}, Katarina}, \au{Luchtenburg, Mark}, \au{King, Rudibert},
  \au{Noack, Bernd} \& \au{Pfeifer, Jens}} \yr{2010} Robust nonlinear control
  versus linear model predictive control of a bluff body wake.  \bt{In {\em 5th
  Flow Control Conference\/}},  \pg{p. 4833}.

\bibitem[Aleksi{\'c}-Roe{\ss}ner {\em et~al.\/}(2014)Aleksi{\'c}-Roe{\ss}ner,
  King, Lehmann, Tadmor \& Morzy{\'n}ski]{king2014}
{\sc \au{Aleksi{\'c}-Roe{\ss}ner, Katarina}, \au{King, Rudibert}, \au{Lehmann,
  Oliver}, \au{Tadmor, Gilead} \& \au{Morzy{\'n}ski, Marek}} \yr{2014}  \at{On
  the need of nonlinear control for efficient model-based wake stabilization}.
  \jt{Theoretical and Computational Fluid Dynamics}  \bvol{28},  \pg{23--49}.

\bibitem[Amestoy {\em et~al.\/}(2000)Amestoy, Duff, L’Excellent \&
  Koster]{mumps}
{\sc \au{Amestoy, Patrick~R}, \au{Duff, Iain~S}, \au{L’Excellent, Jean-Yves}
  \& \au{Koster, Jacko}} \yr{2000} {MUMPS}: a general purpose distributed
  memory sparse solver.  \bt{In {\em International Workshop on Applied Parallel
  Computing\/}},  \pg{pp. 121--130}. Springer.

\bibitem[Apkarian {\em et~al.\/}(2014)Apkarian, Gahinet \&
  Buhr]{apkarian2014multi}
{\sc \au{Apkarian, Pierre}, \au{Gahinet, Pascal} \& \au{Buhr, Craig}} \yr{2014}
  Multi-model, multi-objective tuning of fixed-structure controllers.  \bt{In
  {\em 2014 European Control Conference (ECC)\/}},  \pg{pp. 856--861}. IEEE.

\bibitem[Apkarian \& Noll(2006)]{apkarian2006nonsmooth}
{\sc \au{Apkarian, Pierre} \& \au{Noll, Dominikus}} \yr{2006}  \at{Nonsmooth
  $\mathcal{H}_\infty$ synthesis}.  \jt{IEEE Transactions on Automatic Control}
   \bvol{51}~(1),  \pg{71--86}.

\bibitem[Arbabi {\em et~al.\/}(2018)Arbabi, Korda \& Mezi{\'c}]{arbabi2018}
{\sc \au{Arbabi, Hassan}, \au{Korda, Milan} \& \au{Mezi{\'c}, Igor}} \yr{2018}
  A data-driven {K}oopman model predictive control framework for nonlinear
  partial differential equations.  \bt{In {\em 2018 IEEE Conference on Decision
  and Control (CDC)\/}},  \pg{pp. 6409--6414}. IEEE.

\bibitem[Bagheri(2014)]{bagheri2014}
{\sc \au{Bagheri, Shervin}} \yr{2014}  \at{Effects of weak noise on oscillating
  flows: Linking quality factor, {F}loquet modes, and {K}oopman spectrum}.
  \jt{Physics of Fluids}  \bvol{26}~(9).

\bibitem[Barbagallo {\em et~al.\/}(2009)Barbagallo, Sipp \&
  Schmid]{barbagallo2009}
{\sc \au{Barbagallo, Alexandre}, \au{Sipp, Denis} \& \au{Schmid, Peter~J}}
  \yr{2009}  \at{Closed-loop control of an open cavity flow using reduced-order
  models}.  \jt{Journal of Fluid Mechanics}  \bvol{641},  \pg{1--50}.

\bibitem[Barbagallo {\em et~al.\/}(2011)Barbagallo, Sipp \&
  Schmid]{barbagallo2011input}
{\sc \au{Barbagallo, Alexandre}, \au{Sipp, Denis} \& \au{Schmid, Peter~J}}
  \yr{2011}  \at{Input--output measures for model reduction and closed-loop
  control: application to global modes}.  \jt{Journal of Fluid Mechanics}
  \bvol{685},  \pg{23--53}.

\bibitem[Barkley(2006)]{barkley2006}
{\sc \au{Barkley, Dwight}} \yr{2006}  \at{Linear analysis of the cylinder wake
  mean flow}.  \jt{Europhysics Letters}  \bvol{75}~(5),  \pg{750}.

\bibitem[Barkley {\em et~al.\/}(2000)Barkley, Tuckerman \&
  Golubitsky]{barkley2000bifurcation}
{\sc \au{Barkley, Dwight}, \au{Tuckerman, Laurette~S} \& \au{Golubitsky,
  Martin}} \yr{2000}  \at{Bifurcation theory for three-dimensional flow in the
  wake of a circular cylinder}.  \jt{Physical Review E}  \bvol{61}~(5),
  \pg{5247}.

\bibitem[Bendat \& Piersol(2011)]{bendat2011random}
{\sc \au{Bendat, Julius~S} \& \au{Piersol, Allan~G}} \yr{2011} {\em Random
  data: analysis and measurement procedures\/}.  \publ{John Wiley \& Sons}.

\bibitem[Bengana {\em et~al.\/}(2019)Bengana, Loiseau, Robinet \&
  Tuckerman]{bengana2019}
{\sc \au{Bengana, Y}, \au{Loiseau, J-Ch}, \au{Robinet, J-Ch} \& \au{Tuckerman,
  LS}} \yr{2019}  \at{Bifurcation analysis and frequency prediction in
  shear-driven cavity flow}.  \jt{Journal of Fluid Mechanics}  \bvol{875},
  \pg{725--757}.

\bibitem[Benner {\em et~al.\/}(2022)Benner, Heiland \& Werner]{bennerhinf}
{\sc \au{Benner, Peter}, \au{Heiland, Jan} \& \au{Werner, Steffen~WR}}
  \yr{2022}  \at{Robust output-feedback stabilization for incompressible flows
  using low-dimensional $\mathcal{H}_\infty$-controllers}.  \jt{Computational
  Optimization and Applications}  \bvol{82}~(1),  \pg{225--249}.

\bibitem[Bergmann {\em et~al.\/}(2005)Bergmann, Cordier \&
  Brancher]{bergmann2005}
{\sc \au{Bergmann, Michel}, \au{Cordier, Laurent} \& \au{Brancher,
  Jean-Pierre}} \yr{2005}  \at{Optimal rotary control of the cylinder wake
  using proper orthogonal decomposition reduced-order model}.  \jt{Physics of
  fluids}  \bvol{17}~(9),  \pg{097101}.

\bibitem[Bieker {\em et~al.\/}(2019)Bieker, Peitz, Brunton, Kutz \&
  Dellnitz]{bieker2019}
{\sc \au{Bieker, Katharina}, \au{Peitz, Sebastian}, \au{Brunton, Steven~L},
  \au{Kutz, J~Nathan} \& \au{Dellnitz, Michael}} \yr{2019}  \at{Deep {M}odel
  {P}redictive {C}ontrol with online learning for complex physical systems}.
  \jt{arXiv preprint arXiv:1905.10094} .

\bibitem[Brunton \& Noack(2015)]{brunton2015}
{\sc \au{Brunton, Steven~L} \& \au{Noack, Bernd~R}} \yr{2015}  \at{Closed-loop
  turbulence control: {Progress} and challenges}.  \jt{Applied Mechanics
  Reviews} .

\bibitem[Camarri \& Iollo(2010)]{camarri2010feedback}
{\sc \au{Camarri, Simone} \& \au{Iollo, Angelo}} \yr{2010}  \at{Feedback
  control of the vortex-shedding instability based on sensitivity analysis}.
  \jt{Physics of Fluids}  \bvol{22}~(9),  \pg{094102}.

\bibitem[Carini {\em et~al.\/}(2015)Carini, Pralits \& Luchini]{carini2015}
{\sc \au{Carini, M}, \au{Pralits, JO} \& \au{Luchini, P}} \yr{2015}
  \at{Feedback control of vortex shedding using a full-order optimal
  compensator}.  \jt{Journal of Fluids and Structures}  \bvol{53},
  \pg{15--25}.

\bibitem[Castellanos {\em et~al.\/}(2022)Castellanos, Cornejo~Maceda,
  De~La~Fuente, Noack, Ianiro \& Discetti]{cornejo2022}
{\sc \au{Castellanos, R}, \au{Cornejo~Maceda, GY}, \au{De~La~Fuente, I},
  \au{Noack, BR}, \au{Ianiro, A} \& \au{Discetti, S}} \yr{2022}
  \at{Machine-learning flow control with few sensor feedback and measurement
  noise}.  \jt{Physics of Fluids}  \bvol{34}~(4).

\bibitem[Cheong \& Safonov(2008)]{bumpless_safonov}
{\sc \au{Cheong, Shin-Young} \& \au{Safonov, Michael~G}} \yr{2008}
  \at{Bumpless transfer for adaptive switching controls}.  \jt{IFAC Proceedings
  Volumes}  \bvol{41}~(2),  \pg{14415--14420}.

\bibitem[Dahan {\em et~al.\/}(2012)Dahan, Morgans \& Lardeau]{dahan2012}
{\sc \au{Dahan, Jeremy~A}, \au{Morgans, AS} \& \au{Lardeau, S}} \yr{2012}
  \at{Feedback control for form-drag reduction on a bluff body with a blunt
  trailing edge}.  \jt{Journal of Fluid Mechanics}  \bvol{704},  \pg{360--387}.

\bibitem[Dalla~Longa {\em et~al.\/}(2017)Dalla~Longa, Morgans \&
  Dahan]{dallalonga2017}
{\sc \au{Dalla~Longa, L}, \au{Morgans, AS} \& \au{Dahan, JA}} \yr{2017}
  \at{Reducing the pressure drag of a {D}-shaped bluff body using linear
  feedback control}.  \jt{Theoretical and Computational Fluid Dynamics}
  \bvol{31},  \pg{567--577}.

\bibitem[Demourant \& Poussot-Vassal(2017)]{fabricelmi}
{\sc \au{Demourant, Fabrice} \& \au{Poussot-Vassal, Charles}} \yr{2017}  \at{A
  new frequency-domain subspace algorithm with restricted poles location
  through {LMI} regions and its application to a wind tunnel test}.
  \jt{International Journal of Control}  \bvol{90}~(4),  \pg{779--799}.

\bibitem[Doyle(1978)]{doyle1978guaranteed}
{\sc \au{Doyle, John~C}} \yr{1978}  \at{Guaranteed margins for {LQG}
  regulators}.  \jt{IEEE Transactions on automatic Control}  \bvol{23}~(4),
  \pg{756--757}.

\bibitem[Enns(1984)]{enns1984}
{\sc \au{Enns, Dale~F}} \yr{1984} Model reduction with balanced realizations:
  An error bound and a frequency weighted generalization.  \bt{In {\em The 23rd
  IEEE conference on decision and control\/}},  \pg{pp. 127--132}. IEEE.

\bibitem[Evstafyeva {\em et~al.\/}(2017)Evstafyeva, Morgans \&
  Dalla~Longa]{evstafyeva2017}
{\sc \au{Evstafyeva, O}, \au{Morgans, AS} \& \au{Dalla~Longa, L}} \yr{2017}
  \at{Simulation and feedback control of the {A}hmed body flow exhibiting
  symmetry breaking behaviour}.  \jt{Journal of Fluid Mechanics}  \bvol{817},
  \pg{R2}.

\bibitem[Flinois \& Morgans(2016)]{flinois2016feedback}
{\sc \au{Flinois, Thibault~LB} \& \au{Morgans, Aimee~S}} \yr{2016}
  \at{Feedback control of unstable flows: a direct modelling approach using the
  {E}igensystem {R}ealisation {A}lgorithm}.  \jt{Journal of Fluid Mechanics}
  \bvol{793},  \pg{41--78}.

\bibitem[Garnier {\em et~al.\/}(2021)Garnier, Viquerat, Rabault, Larcher,
  Kuhnle \& Hachem]{viqueratreview}
{\sc \au{Garnier, Paul}, \au{Viquerat, Jonathan}, \au{Rabault, Jean},
  \au{Larcher, Aur{\'e}lien}, \au{Kuhnle, Alexander} \& \au{Hachem, Elie}}
  \yr{2021}  \at{A review on {D}eep {R}einforcement {L}earning for fluid
  mechanics}.  \jt{Computers \& Fluids}  \bvol{225},  \pg{104973}.

\bibitem[Gerhard {\em et~al.\/}(2003)Gerhard, Pastoor, King, Noack, Dillmann,
  Morzynski \& Tadmor]{king2003}
{\sc \au{Gerhard, Johannes}, \au{Pastoor, Mark}, \au{King, Rudibert},
  \au{Noack, Bernd}, \au{Dillmann, Andreas}, \au{Morzynski, Marek} \&
  \au{Tadmor, Gilead}} \yr{2003} Model-based control of vortex shedding using
  low-dimensional {G}alerkin models.  \bt{In {\em 33rd AIAA fluid dynamics
  conference and exhibit\/}},  \pg{p. 4262}.

\bibitem[Ghraieb {\em et~al.\/}(2021)Ghraieb, Viquerat, Larcher, Meliga \&
  Hachem]{ghraieb2021}
{\sc \au{Ghraieb, Hassan}, \au{Viquerat, Jonathan}, \au{Larcher, Aur{\'e}lien},
  \au{Meliga, Philippe} \& \au{Hachem, Elie}} \yr{2021}  \at{Single-step deep
  reinforcement learning for open-loop control of laminar and turbulent flows}.
   \jt{Physical Review Fluids}  \bvol{6}~(5),  \pg{053902}.

\bibitem[Gugercin \& Antoulas(2004)]{balrealsurvey}
{\sc \au{Gugercin, Serkan} \& \au{Antoulas, Athanasios~C}} \yr{2004}  \at{A
  survey of model reduction by balanced truncation and some new results}.
  \jt{International Journal of Control}  \bvol{77}~(8),  \pg{748--766}.

\bibitem[Gustavsen \& Semlyen(1999)]{vectorfitting}
{\sc \au{Gustavsen, Bjorn} \& \au{Semlyen, Adam}} \yr{1999}  \at{Rational
  approximation of frequency domain responses by vector fitting}.  \jt{IEEE
  Transactions on power delivery}  \bvol{14}~(3),  \pg{1052--1061}.

\bibitem[Heiland \& Werner(2023)]{heiland2023}
{\sc \au{Heiland, Jan} \& \au{Werner, Steffen~WR}} \yr{2023}
  \at{Low-complexity linear parameter-varying approximations of incompressible
  {N}avier-{S}tokes equations for truncated state-dependent {R}iccati
  feedback}.  \jt{arXiv preprint arXiv:2303.11515} .

\bibitem[Huang {\em et~al.\/}(2017)Huang, Jin, Lasagna, Chernyshenko \&
  Tutty]{lasagna2017}
{\sc \au{Huang, Deqing}, \au{Jin, Bo}, \au{Lasagna, Davide}, \au{Chernyshenko,
  Sergei} \& \au{Tutty, Owen}} \yr{2017}  \at{Expensive control of long-time
  averages using sum of squares and its application to a laminar wake flow}.
  \jt{IEEE Transactions on Control Systems Technology}  \bvol{25}~(6),
  \pg{2073--2086}.

\bibitem[Illingworth(2016)]{illingworth2016}
{\sc \au{Illingworth, Simon~J}} \yr{2016}  \at{Model-based control of vortex
  shedding at low {R}eynolds numbers}.  \jt{Theoretical and Computational Fluid
  Dynamics}  \bvol{30}~(5),  \pg{429--448}.

\bibitem[Illingworth {\em et~al.\/}(2011)Illingworth, Morgans \&
  Rowley]{illingworth2011}
{\sc \au{Illingworth, Simon~J}, \au{Morgans, Aimee~S} \& \au{Rowley,
  Clarence~W}} \yr{2011}  \at{Feedback control of flow resonances using
  balanced reduced-order models}.  \jt{Journal of Sound and Vibration}
  \bvol{330}~(8),  \pg{1567--1581}.

\bibitem[Illingworth {\em et~al.\/}(2012)Illingworth, Morgans \&
  Rowley]{illingworth2012}
{\sc \au{Illingworth, Simon~J}, \au{Morgans, Aimee~S} \& \au{Rowley,
  Clarence~W}} \yr{2012}  \at{Feedback control of cavity flow oscillations
  using simple linear models}.  \jt{Journal of Fluid Mechanics}  \bvol{709},
  \pg{223--248}.

\bibitem[Jin {\em et~al.\/}(2020)Jin, Illingworth \& Sandberg]{jin2020feedback}
{\sc \au{Jin, Bo}, \au{Illingworth, Simon~J} \& \au{Sandberg, Richard~D}}
  \yr{2020}  \at{Feedback control of vortex shedding using a resolvent-based
  modelling approach}.  \jt{Journal of Fluid Mechanics}  \bvol{897}.

\bibitem[Juang \& Suzuki(1988)]{ERAFD}
{\sc \au{Juang, J-N} \& \au{Suzuki, Hideto}} \yr{1988}  \at{An {E}igensystem
  {R}ealization {A}lgorithm in {F}requency {D}omain for modal parameter
  identification}.  \jt{Journal of Vibration, Acoustics, Stress and Reliability
  in Design}  \bvol{110}~(1),  \pg{24}.

\bibitem[Jussiau {\em et~al.\/}(2022)Jussiau, Leclercq, Demourant \&
  Apkarian]{jussiau2022learning}
{\sc \au{Jussiau, William}, \au{Leclercq, Colin}, \au{Demourant, Fabrice} \&
  \au{Apkarian, Pierre}} \yr{2022}  \at{Learning linear feedback controllers
  for suppressing the vortex-shedding flow past a cylinder}.  \jt{IEEE Control
  Systems Letters}  \bvol{6},  \pg{3212--3217}.

\bibitem[Kim \& Bewley(2007)]{bewleylinear}
{\sc \au{Kim, John} \& \au{Bewley, Thomas~R}} \yr{2007}  \at{A linear systems
  approach to flow control}.  \jt{Annu. Rev. Fluid Mech.}  \bvol{39},
  \pg{383--417}.

\bibitem[King {\em et~al.\/}(2005)King, Seibold, Lehmann, Noack, Morzy{\'n}ski
  \& Tadmor]{king2005}
{\sc \au{King, Rudibert}, \au{Seibold, Meline}, \au{Lehmann, Oliver},
  \au{Noack, Bernd~R}, \au{Morzy{\'n}ski, Marek} \& \au{Tadmor, Gilead}}
  \yr{2005}  \at{Nonlinear flow control based on a low dimensional model of
  fluid flow}.  \jt{Control and Observer Design for Nonlinear Finite and
  Infinite Dimensional Systems}  \pg{pp. 369--386}.

\bibitem[Korda \& Mezi{\'c}(2018{\natexlab{{\em a\/}}})]{kordamezic}
{\sc \au{Korda, Milan} \& \au{Mezi{\'c}, Igor}} \yr{2018{\natexlab{{\em a\/}}}}
   \at{Linear predictors for nonlinear dynamical systems: Koopman operator
  meets model predictive control}.  \jt{Automatica}  \bvol{93},  \pg{149--160}.

\bibitem[Korda \& Mezi{\'c}(2018{\natexlab{{\em b\/}}})]{kordamezicedmd}
{\sc \au{Korda, Milan} \& \au{Mezi{\'c}, Igor}} \yr{2018{\natexlab{{\em b\/}}}}
   \at{On convergence of extended dynamic mode decomposition to the koopman
  operator}.  \jt{Journal of Nonlinear Science}  \bvol{28},  \pg{687--710}.

\bibitem[Lasagna {\em et~al.\/}(2016)Lasagna, Huang, Tutty \&
  Chernyshenko]{lasagna2016}
{\sc \au{Lasagna, Davide}, \au{Huang, Deqing}, \au{Tutty, Owen~R} \&
  \au{Chernyshenko, Sergei}} \yr{2016}  \at{Sum-of-squares approach to feedback
  control of laminar wake flows}.  \jt{Journal of Fluid Mechanics}  \bvol{809},
   \pg{628--663}.

\bibitem[Laub {\em et~al.\/}(1987)Laub, Heath, Paige \& Ward]{balreal1}
{\sc \au{Laub, Alanj}, \au{Heath, MICHAELT}, \au{Paige, C} \& \au{Ward, R}}
  \yr{1987}  \at{Computation of system balancing transformations and other
  applications of simultaneous diagonalization algorithms}.  \jt{IEEE
  Transactions on Automatic Control}  \bvol{32}~(2),  \pg{115--122}.

\bibitem[Leclercq {\em et~al.\/}(2019)Leclercq, Demourant, Poussot-Vassal \&
  Sipp]{leclercq2019}
{\sc \au{Leclercq, Colin}, \au{Demourant, Fabrice}, \au{Poussot-Vassal,
  Charles} \& \au{Sipp, Denis}} \yr{2019}  \at{Linear iterative method for
  closed-loop control of quasiperiodic flows}.  \jt{Journal of Fluid Mechanics}
   \bvol{868},  \pg{26--65}.

\bibitem[Leclercq \& Sipp(2023)]{leclercq2023}
{\sc \au{Leclercq, Colin} \& \au{Sipp, Denis}} \yr{2023}  \at{Mean resolvent
  operator of a statistically steady flow}.  \jt{Journal of Fluid Mechanics}
  \bvol{968},  \pg{A13}.

\bibitem[Lehtomaki {\em et~al.\/}(1981)Lehtomaki, Sandell \&
  Athans]{lqr_robustness}
{\sc \au{Lehtomaki, N.}, \au{Sandell, N.} \& \au{Athans, M.}} \yr{1981}
  \at{Robustness results in linear-quadratic {G}aussian based multivariable
  control designs}.  \jt{IEEE Transactions on Automatic Control}
  \bvol{26}~(1),  \pg{75--93}.

\bibitem[Li \& Morgans(2016)]{li2016feedback}
{\sc \au{Li, Jingxuan} \& \au{Morgans, Aimee~S}} \yr{2016}  \at{Feedback
  control of combustion instabilities from within limit cycle oscillations
  using hinfinity loop-shaping and the nu-gap metric}.  \jt{Proceedings of the
  Royal Society A: Mathematical, Physical and Engineering Sciences}
  \bvol{472}~(2191),  \pg{20150821}.

\bibitem[Liu {\em et~al.\/}(1994)Liu, Jacques \& Miller]{forse}
{\sc \au{Liu, K}, \au{Jacques, RN} \& \au{Miller, DW}} \yr{1994} {F}requency
  domain structural system identification by {O}bservability {R}ange {S}pace
  {E}xtraction.  \bt{In {\em Proceedings of 1994 American Control
  Conference-ACC'94\/}}, ,  \vol{vol.~1},  \pg{pp. 107--111}. IEEE.

\bibitem[Liu {\em et~al.\/}(2018)Liu, Sun, Cattafesta, Ukeiley \&
  Taira]{liu2018resolvent}
{\sc \au{Liu, Qiong}, \au{Sun, Yiyang}, \au{Cattafesta, Louis~N}, \au{Ukeiley,
  Lawrence~S} \& \au{Taira, Kunihiko}} \yr{2018} Resolvent analysis of
  compressible flow over a long rectangular cavity.  \bt{In {\em 2018 AIAA
  Aerospace Sciences Meeting\/}},  \pg{p. 0588}.

\bibitem[Logg {\em et~al.\/}(2012)Logg, Mardal \& Wells]{fenics}
{\sc \au{Logg, Anders}, \au{Mardal, Kent-Andre} \& \au{Wells, Garth}} \yr{2012}
  {\em Automated solution of differential equations by the finite element
  method: The {FEniCS} book\/}, ,  \vol{vol.~84}.  \publ{Springer Science \&
  Business Media}.

\bibitem[Maceda {\em et~al.\/}(2021)Maceda, Li, Lusseyran, Morzy{\'n}ski \&
  Noack]{cornejo2021}
{\sc \au{Maceda, Guy Y~Cornejo}, \au{Li, Yiqing}, \au{Lusseyran,
  Fran{\c{c}}ois}, \au{Morzy{\'n}ski, Marek} \& \au{Noack, Bernd~R}} \yr{2021}
  \at{Stabilization of the fluidic pinball with gradient-enriched machine
  learning control}.  \jt{Journal of Fluid Mechanics}  \bvol{917},  \pg{A42}.

\bibitem[Moore(1981)]{moore1981}
{\sc \au{Moore, Bruce}} \yr{1981}  \at{{P}rincipal {C}omponent {A}nalysis in
  linear systems: Controllability, observability, and model reduction}.
  \jt{IEEE transactions on automatic control}  \bvol{26}~(1),  \pg{17--32}.

\bibitem[Morton {\em et~al.\/}(2018)Morton, Jameson, Kochenderfer \&
  Witherden]{morton2018}
{\sc \au{Morton, Jeremy}, \au{Jameson, Antony}, \au{Kochenderfer, Mykel~J} \&
  \au{Witherden, Freddie}} \yr{2018}  \at{Deep dynamical modeling and control
  of unsteady fluid flows}.  \jt{Advances in Neural Information Processing
  Systems}  \bvol{31}.

\bibitem[Otto {\em et~al.\/}(2022)Otto, Peitz \& Rowley]{otto2022}
{\sc \au{Otto, Samuel~E}, \au{Peitz, Sebastian} \& \au{Rowley, Clarence~W}}
  \yr{2022}  \at{Learning bilinear models of actuated {K}oopman generators from
  partially-observed trajectories}.  \jt{arXiv preprint arXiv:2209.09977} .

\bibitem[Ozdemir \& Gumussoy(2017)]{vectorfitting2}
{\sc \au{Ozdemir, Ahmet~Arda} \& \au{Gumussoy, Suat}} \yr{2017}  \at{Transfer
  function estimation in system identification toolbox via vector fitting}.
  \jt{IFAC-PapersOnLine}  \bvol{50}~(1),  \pg{6232--6237}.

\bibitem[Page \& Kerswell(2019)]{kerswell}
{\sc \au{Page, Jacob} \& \au{Kerswell, Rich~R}} \yr{2019}  \at{Koopman mode
  expansions between simple invariant solutions}.  \jt{Journal of Fluid
  Mechanics}  \bvol{879},  \pg{1--27}.

\bibitem[Paris {\em et~al.\/}(2021)Paris, Beneddine \& Dandois]{paris}
{\sc \au{Paris, Romain}, \au{Beneddine, Samir} \& \au{Dandois, Julien}}
  \yr{2021}  \at{Robust flow control and optimal sensor placement using deep
  reinforcement learning}.  \jt{Journal of Fluid Mechanics}  \bvol{913}.

\bibitem[Park {\em et~al.\/}(1994)Park, Ladd \& Hendricks]{park1994}
{\sc \au{Park, DS}, \au{Ladd, DM} \& \au{Hendricks, EW}} \yr{1994}
  \at{Feedback control of von {K}{\'a}rm{\'a}n vortex shedding behind a
  circular cylinder at low {R}eynolds numbers}.  \jt{Physics of fluids}
  \bvol{6}~(7),  \pg{2390--2405}.

\bibitem[Paxman(2004)]{paxmanphd}
{\sc \au{Paxman, Jonathan~Patrick}} \yr{2004}  \at{Switching controllers:
  Realization, initialization and stability}. PhD thesis, University of
  Cambridge.

\bibitem[Peitz \& Klus(2019)]{peitz2019switch}
{\sc \au{Peitz, Sebastian} \& \au{Klus, Stefan}} \yr{2019}  \at{{K}oopman
  operator-based model reduction for switched-system control of {PDE}s}.
  \jt{Automatica}  \bvol{106},  \pg{184--191}.

\bibitem[Peitz \& Klus(2020)]{peitz2020summary}
{\sc \au{Peitz, Sebastian} \& \au{Klus, Stefan}} \yr{2020}  \at{Feedback
  control of nonlinear {PDE}s using data-efficient reduced order models based
  on the {K}oopman operator}.  \jt{The {K}oopman Operator in Systems and
  Control: Concepts, Methodologies, and Applications}  \pg{pp. 257--282}.

\bibitem[Peitz {\em et~al.\/}(2020)Peitz, Otto \& Rowley]{peitz2020interp}
{\sc \au{Peitz, Sebastian}, \au{Otto, Samuel~E} \& \au{Rowley, Clarence~W}}
  \yr{2020}  \at{Data-driven model predictive control using interpolated
  {K}oopman generators}.  \jt{SIAM Journal on Applied Dynamical Systems}
  \bvol{19}~(3),  \pg{2162--2193}.

\bibitem[Pernebo \& Silverman(1982)]{pernebo1982}
{\sc \au{Pernebo, Lars} \& \au{Silverman, Leonard}} \yr{1982}  \at{Model
  reduction via balanced state space representations}.  \jt{IEEE Transactions
  on Automatic Control}  \bvol{27}~(2),  \pg{382--387}.

\bibitem[Proctor {\em et~al.\/}(2016)Proctor, Brunton \& Kutz]{dmdc}
{\sc \au{Proctor, Joshua~L}, \au{Brunton, Steven~L} \& \au{Kutz, J~Nathan}}
  \yr{2016}  \at{Dynamic mode decomposition with control}.  \jt{SIAM Journal on
  Applied Dynamical Systems}  \bvol{15}~(1),  \pg{142--161}.

\bibitem[Rabault {\em et~al.\/}(2019)Rabault, Kuchta, Jensen, R{\'e}glade \&
  Cerardi]{rabault}
{\sc \au{Rabault, Jean}, \au{Kuchta, Miroslav}, \au{Jensen, Atle},
  \au{R{\'e}glade, Ulysse} \& \au{Cerardi, Nicolas}} \yr{2019}  \at{Artificial
  neural networks trained through {D}eep {R}einforcement {L}earning discover
  control strategies for active flow control}.  \jt{Journal of fluid mechanics}
   \bvol{865},  \pg{281--302}.

\bibitem[Rowley(2005)]{rowley2005model}
{\sc \au{Rowley, Clarence~W}} \yr{2005}  \at{Model reduction for fluids, using
  balanced proper orthogonal decomposition}.  \jt{International Journal of
  Bifurcation and Chaos}  \bvol{15}~(03),  \pg{997--1013}.

\bibitem[Rowley \& Juttijudata(2005)]{rowley2005jutti}
{\sc \au{Rowley, Clarance~W} \& \au{Juttijudata, Vejapong}} \yr{2005}
  Model-based control and estimation of cavity flow oscillations.  \bt{In {\em
  Proceedings of the 44th IEEE Conference on Decision and Control\/}},  \pg{pp.
  512--517}. IEEE.

\bibitem[Schmid \& Sipp(2016)]{schmid2016}
{\sc \au{Schmid, Peter~J} \& \au{Sipp, Denis}} \yr{2016}  \at{Linear control of
  oscillator and amplifier flows}.  \jt{Physical Review Fluids}  \bvol{1}~(4),
  \pg{040501}.

\bibitem[Schoukens {\em et~al.\/}(1991)Schoukens, Guillaume \&
  Pintelon]{schoukens1991}
{\sc \au{Schoukens, Joannes}, \au{Guillaume, Patrick} \& \au{Pintelon, Rik}}
  \yr{1991} Design of multisine excitations.  \bt{In {\em International
  Conference on Control 1991. Control'91\/}},  \pg{pp. 638--643}. IET.

\bibitem[Schoukens {\em et~al.\/}(2008)Schoukens, Lataire, Pintelon \&
  Vandersteen]{schoukens2008robustness}
{\sc \au{Schoukens, Joannes}, \au{Lataire, John}, \au{Pintelon, R} \&
  \au{Vandersteen, G}} \yr{2008} Robustness issues of the equivalent linear
  representation of a nonlinear system.  \bt{In {\em 2008 IEEE instrumentation
  and measurement technology conference\/}},  \pg{pp. 332--335}. IEEE.

\bibitem[Schoukens {\em et~al.\/}(2016)Schoukens, Vaes \&
  Pintelon]{schoukens2016linear}
{\sc \au{Schoukens, Johan}, \au{Vaes, Mark} \& \au{Pintelon, Rik}} \yr{2016}
  \at{Linear system identification in a nonlinear setting: Nonparametric
  analysis of the nonlinear distortions and their impact on the best linear
  approximation}.  \jt{IEEE Control Systems Magazine}  \bvol{36}~(3),
  \pg{38--69}.

\bibitem[Sipp \& Schmid(2016)]{sipp2016}
{\sc \au{Sipp, Denis} \& \au{Schmid, Peter~J}} \yr{2016}  \at{Linear
  closed-loop control of fluid instabilities and noise-induced perturbations: a
  review of approaches and tools}.  \jt{Applied Mechanics Reviews}
  \bvol{68}~(2).

\bibitem[Son \& Choi(2018)]{choi2018}
{\sc \au{Son, Donggun} \& \au{Choi, Haecheon}} \yr{2018}  \at{Iterative
  feedback tuning of the proportional-integral-differential control of flow
  over a circular cylinder}.  \jt{IEEE Transactions on Control Systems
  Technology}  \bvol{27}~(4),  \pg{1385--1396}.

\bibitem[Syrmos {\em et~al.\/}(1997)Syrmos, Abdallah, Dorato \&
  Grigoriadis]{staticoutputfeedback}
{\sc \au{Syrmos, Vassilis~L}, \au{Abdallah, Chaouki~T}, \au{Dorato, Peter} \&
  \au{Grigoriadis, Karolos}} \yr{1997}  \at{Static output feedback —- a
  survey}.  \jt{Automatica}  \bvol{33}~(2),  \pg{125--137}.

\bibitem[Van~Overschee \& De~Moor(2012)]{van2012subspace}
{\sc \au{Van~Overschee, Peter} \& \au{De~Moor, BL0888}} \yr{2012} {\em Subspace
  identification for linear systems: Theory—Implementation—Applications\/}.
   \publ{Springer Science \& Business Media}.

\bibitem[Viquerat {\em et~al.\/}(2022)Viquerat, Meliga, Larcher \&
  Hachem]{viqueratreviewupdate}
{\sc \au{Viquerat, Jonathan}, \au{Meliga, Philippe}, \au{Larcher, A} \&
  \au{Hachem, E}} \yr{2022}  \at{A review on deep reinforcement learning for
  fluid mechanics: An update}.  \jt{Physics of Fluids}  \bvol{34}~(11).

\bibitem[Weller {\em et~al.\/}(2009)Weller, Camarri \&
  Iollo]{weller2009feedback}
{\sc \au{Weller, Jessie}, \au{Camarri, Simone} \& \au{Iollo, Angelo}} \yr{2009}
   \at{Feedback control by low-order modelling of the laminar flow past a bluff
  body}.  \jt{Journal of fluid mechanics}  \bvol{634},  \pg{405--418}.

\bibitem[Williams {\em et~al.\/}(2015)Williams, Kevrekidis \& Rowley]{edmd}
{\sc \au{Williams, Matthew~O}, \au{Kevrekidis, Ioannis~G} \& \au{Rowley,
  Clarence~W}} \yr{2015}  \at{A data--driven approximation of the koopman
  operator: Extending dynamic mode decomposition}.  \jt{Journal of Nonlinear
  Science}  \bvol{25},  \pg{1307--1346}.

\bibitem[Xia {\em et~al.\/}(2023)Xia, Zhang, Kerrigan \& Rigas]{rigas2023}
{\sc \au{Xia, Chengwei}, \au{Zhang, Junjie}, \au{Kerrigan, Eric~C} \&
  \au{Rigas, Georgios}} \yr{2023}  \at{Active flow control for bluff body drag
  reduction using reinforcement learning with partial measurements}.  \jt{arXiv
  preprint arXiv:2307.12650} .

\bibitem[Yun \& Lee(2022)]{yun2022}
{\sc \au{Yun, Jinhyeok} \& \au{Lee, Jungil}} \yr{2022}  \at{Active proportional
  feedback control of turbulent flow over a circular cylinder with averaged
  velocity sensor}.  \jt{Physics of Fluids}  \bvol{34}~(9).

\bibitem[Zaccarian \& Teel(2005)]{bumpless_zaccarian}
{\sc \au{Zaccarian, Luca} \& \au{Teel, Andrew~R}} \yr{2005}  \at{The {L2}
  bumpless transfer problem for linear plants: Its definition and solution}.
  \jt{Automatica}  \bvol{41}~(7),  \pg{1273--1280}.

\bibitem[Zhou \& Doyle(1998)]{zhoudoyle}
{\sc \au{Zhou, Kemin} \& \au{Doyle, John~Comstock}} \yr{1998} {\em Essentials
  of robust control\/}, ,  \vol{vol. 104}.  \publ{Prentice Hall Upper Saddle
  River, NJ}.

\bibitem[Zhou {\em et~al.\/}(1999)Zhou, Salomon \& Wu]{zhou1999}
{\sc \au{Zhou, Kemin}, \au{Salomon, Gregory} \& \au{Wu, Eva}} \yr{1999}
  \at{Balanced realization and model reduction for unstable systems}.
  \jt{International Journal of Robust and Nonlinear Control: IFAC-Affiliated
  Journal}  \bvol{9}~(3),  \pg{183--198}.

\end{thebibliography}
%%%%%%%%%%%%%%%%%%%%%%%%%%%%%%%%%%%%%%

\end{document}